\documentclass[12pt]{article}
\pdfoutput=1
\usepackage{latexsym, graphicx,cite}
\usepackage{amsmath}
\usepackage{amssymb}
\usepackage{amsthm}
\usepackage{tkz-graph}
\usepackage{graphicx}
\usepackage{tikz}
\usepackage{subcaption}
\usetikzlibrary{positioning}

\usepackage[top = 1. in,bottom = 0.98 in, left = 1 in, right=1 in]{geometry}

 \newcommand{\arXiv}[1]{\href{http://www.arXiv.org/abs/#1}{arXiv:#1}}
\usepackage[colorlinks=true, linkcolor=blue, bookmarks=true]{hyperref}

\makeatletter
\renewcommand\section{\@startsection {section}{1}{\z@}%
                  {-3.5ex \@plus -1ex \@minus -.2ex}
                  {2.3ex \@plus.2ex}%
                  {\normalfont\large\bfseries}}
\renewcommand\subsection{\@startsection{subsection}{2}{\z@}%
                   {-3.25ex\@plus -1ex \@minus -.2ex}%
                   {1.5ex \@plus .2ex}%
                   {\normalfont\bfseries}}
\makeatother

\newcommand{\beq}{\begin{equation}}
\newcommand{\eeq}{\end{equation}}
\newcommand{\beqa}{\begin{eqnarray}}
\newcommand{\eeqa}{\end{eqnarray}}
\newcommand{\ber}{\begin{array}}
\newcommand{\eer}{\end{array}}
\newcommand{\dsty}{\displaystyle}
\newcommand{\s}{\sigma}
\newcommand{\te}{\theta}
\newcommand{\de}{\delta}
\newcommand{\al}{\alpha}
\newcommand{\be}{\beta}
\newcommand{\iu}{\mathrm{i}\mkern1mu}
\DeclareMathOperator{\Tr}{Tr}

\begin{document}
\begin{titlepage}
\begin{flushright}
\phantom{arXiv:yymm.nnnn}
\end{flushright}
\vspace{-5mm}
\begin{center}
{\huge\bf The birth of geometry in\vspace{2mm}\\ exponential random graphs}\\
\vskip 15mm
{\large Pawat Akara-pipattana$^{a}$, Thiparat Chotibut$^{a,b}$ and Oleg Evnin$^{a,c}$}
\vskip 7mm
{\em $^a$ Department of Physics, Faculty of Science, Chulalongkorn University,
Bangkok, Thailand}
\vskip 3mm
{\em $^b$ Chula Intelligent and Complex Systems, Faculty of Science,\\
Chulalongkorn University, Bangkok, Thailand}
\vskip 3mm
{\em $^c$ Theoretische Natuurkunde, Vrije Universiteit Brussel and\\
The International Solvay Institutes, Brussels, Belgium}
\vskip 7mm
{\small\noindent {\tt akarapawat@gmail.com, thiparatc@gmail.com, oleg.evnin@gmail.com}}
\vskip 20mm
\end{center}
\begin{center}
{\bf ABSTRACT}\vspace{3mm}
\end{center}
Inspired by the prospect of having discretized spaces emerge from random graphs, we construct a collection 
of simple and explicit exponential random graph models that enjoy, in an appropriate parameter regime, a roughly constant vertex degree and 
form very large numbers of simple polygons (triangles or squares). The models avoid the collapse phenomena that
plague naive graph Hamiltonians based on triangle or square counts. More than that, statistically significant numbers
of other geometric primitives (small pieces of regular lattices, cubes) emerge in our ensemble, even though they are not 
in any way explicitly pre-programmed into the formulation of the graph Hamiltonian, which only depends on properties
of paths of length 2. While much of our motivation comes from hopes to construct a graph-based theory
of random geometry (Euclidean quantum gravity), our presentation is completely self-contained within the context
of exponential random graph theory, and the range of potential applications is considerably more broad.

\vfill

\end{titlepage}


\section{Introduction}

The idea that all of physics should be a manifestation of geometry has been at the heart of the theory of fundamental interactions over the last hundred  years. But can geometry itself arise from something that is more fundamental? In particular, starting with nothing but a set, can relations between its elements serve as precursors of geometric structures?\footnote{A notable research program that shares much of the same spirit, not directly related to our current pursuits, is the theory of causal sets \cite{cs}. More broadly, considerable advocacy for the emergence of all laws of physics from operations on discrete sets
has been put forward in \cite{W1,W2}.}

One concrete realization of such abstract ideas is an attempt to derive discretized spaces from ensembles of random graphs.
Extensive studies have been undertaken of random discretized spaces (simplicial complexes), with the aim of obtaining a Euclidean theory of 
quantum gravity \cite{QG,CDT,DGR}. In recent works \cite{Trugenberger,KT,KTB}, an approach was formulated to deriving similar structures from an ensemble of random graphs (related earlier work can be found in \cite{ising2st}).
The edges of simplicial complexes, after all, can be viewed as extremely constrained graphs.
The random graphs considered in \cite{Trugenberger,KT,KTB} are themselves rather constrained. For example, in \cite{KT}, one deals with bipartite, $2d$-regular
graphs satisfying the hard-core constraint (no two squares may share more than one edge), while in \cite{KTB} only the bipartiteness condition is dropped. Such constraints have to be rigidly imposed at each vertex and each edge. 

We would like to develop less constrained models of appeal for studying the emergence of geometry. A context that we find particularly attractive is exponential random graph models (for textbook treatments,
see \cite{newman,CAR}). These models, which are natural from both physical and information-theoretic viewpoints, assign Boltzmann-Gibbs exponential weights to all possible ways of connecing a set of $N$ points to each other, without {\it a priori} excluding any specific graphs. In accordance with the principle of maximum entropy \cite{jaynes}, this  probability measure maximizes the Shannon entropy (the lack of knowledge) subject to the constraints on the expectation values of selected observables, say, the total number of edges, triangles, etc. 

In the context of emergent geometry, one immediate question is then as follows: can one formulate an exponential random graph model that gives rise to sparse graphs with small degree variation, and a large number of geometric primitives (say, triangles or squares, but we shall also look at more extended geometric structures)? The focus on sparseness (fixed mean degree in the large graph limit) and small degree variation comes from broadly aiming at generating graphs that resemble simplicial complexes (and is similar to the considerations of \cite{Trugenberger,KT,KTB}). Producing large numbers of geometric primitives, on the other hand, is known to be nontrivial: already the classic work \cite{strauss} notes that attempting to specify 
the expectation value of the number of triangles in simple exponential graph models leads to the emergence of very sparse (nearly empty) and very dense (nearly complete) graph configurations, rather than to `reasonable' graphs with finite degrees and large numbers of triangles. A similar undesirable graph collapse phenomenon occurs in other exponential graph models whose Hamiltonians favor high numbers of specific motifs; see, for example, a description of
such phase transitions between extremely dense and extremely sparse phases in a simple two-star exponential random graph model in \cite{BPG}.
Some solutions to this graph collapse problem have been proposed \cite{KT,KTB,Kr}, but they involve controlling the connectivities of each vertex individually, and such rigid constraints are what we would like to avoid. Another way to escape the collapse has been proposed in \cite{newexp}, which is in the form of a conventional
exponential graph model, but with a complicated graph Hamiltonian, and not adapted to generation of graphs with approximately constant degrees. Some further related literature on random graphs viewed from a statistical physics perspective can be found in \cite{BJK,PN1,PN2,AC,GV,corr}; the problem has also attracted attention of mathematicians working in probability theory \cite{math1,math2,math3}.

We shall then formulate and study two closely related models that meet the above guidelines. In order to attain an approximately constant vertex degree in the context of an exponential random graph model, we shall introduce a large negative two-star term in the graph Hamiltonian (this term can also be seen as the total number of paths of length 2). As known since \cite{AC}, such terms serve to reduce degree variance.
They do not, however, produce large numbers of geometric primitives, and our main objective is to design novel extra terms in the graph Hamiltonian that attain this objective. These terms can be constructed in two flavors, either to boost the number of triangles or the number of squares (and the choice of one of these two options defines the two closely related models we construct).
For reasons explained above, naive triangle or square counts would not have worked in this capacity, as they lead to graph collapse. We take inspiration from the hard-core constraint used in \cite{KT,KTB} and replace this rigid constraint by a term in the graph Hamiltonian that favors configurations that meet the hard-core condition. We introduce the new models explicitly in section \ref{sec: models}. Numerical results from Monte Carlo sampling of our graph ensembles are reported in section \ref{sec: numerical}, justifying our expectations for the formation of simple geometric structures. Besides the triangle and squares, we also observe small pieces of triangular lattices and cubes. These latter structures are emergent, as they do not explicitly appear in any way in the graph Hamiltonian, which is entirely formulated in terms of properties of paths of length 2. We conclude with brief comments on the analytic properties of our models in section \ref{sec: analytic}, possible emergence of large-scale geometric features (as opposed to the small-scale geometric primitives whose formation we see explicitly) in section \ref{sec: large_geo}, and a discussion of the prospects in section \ref{sec: conclusion}.


\section{Definition of the models}\label{sec: models}

It is customary in exponential random graph models \cite{newman,CAR} to represent graphs as a set or random connections between $N$
points (vertices) labelled by $i=1, \dots, N$. The connections are encoded in the {\it adjacency matrix} whose element $c_{ij}$ equal 1 if there is an edge between vertices $i$ and $j$ and zero otherwise. For undirected graphs without self-looping, $c$ is a symmetric matrix with zero diagonal elements. One then defines a {\it graph Hamiltonian} $H$ as a function of $c$ and assigns Boltzmann weights
proportional to $e^{-H(c)}$ to each graph configuration. The expectation value of an observable $A(c)$ in this ensemble is then given by
\beq
\langle A\rangle = \frac1Z \sum_{\{c\}} A(c) e^{-H(c)},
\label{expect}
\eeq
where the sum is over all possible adjacency matrices and $Z\equiv  \sum_{\{c\}} e^{-H(c)}$ is the partition function. In practice, as will be specified in the next section, one would
use Monte Carlo simulations to sample this ensemble.  Typically, the graph Hamiltonian is of the form $\sum_k \alpha_k F_k(c)$, where the Lagrange multipliers $\al_k$ are thermodynamic parameters that control the equilibruim expectation values of quantities $F_k$.

Exponential random graph models have been widely considered in the literature, but typically from perspectives of social science \cite{social}, network theory \cite{newman,CAR} or statistical physics \cite{BJK,PN1,PN2,AC,GV,corr}. To the best of our knowledge, our treatment is the first instance
where completely conventional exponential random graph models (with a finite number of Lagrange multipliers, and without hard constraints on the type of graphs included in the ensemble) are evoked
in relation to random geometry topics. Closely related past studies exist \cite{Trugenberger, KT,KTB}
where the setup is very similar to exponential random graph models, but only graphs satisfying certain constraints are included in the ensemble.

Before we can proceed, we should decide on the choice of the graph Hamiltonian. A good starting point is the two-star model that has been extensively studied in the literature \cite{newman,CAR,PN1,BJK,AC}:
\beq
H_{2*}=-\al \sum_{ij} c_{ij} -\be \sum_{ijk} c_{ij}c_{jk}.
\label{2st}
\eeq
In the two-star model, the Lagrange multiplier $\alpha$ controls the number of edges in the graph, while the Lagrange multiplier $\be$ controls the number of `two-stars', which is related\footnote{Our conventions differ from some of the literature, but are equivalent to it, and convenient for our purposes. The actual number of edges is $\sum_{i>j} c_{ij}$, and the number of two-stars is $\sum_{i>k}\sum_j c_{ij}c_{jk}$, but the Lagrangian multipliers for these quantities are related to $\al$ and $\be$ in (\ref{2st}) by a simple linear redefinition.}  to the total number of paths of length 2. Our first aim is to obtain graphs with a small ($N$-independent) roughly constant vertex degree, in other words, graphs that are sparse and (approximately) regular. We can use $\alpha$ to control the vertex degree, while, as explained in \cite{AC}, a large negative $\be$ can suppress degree variance. This effect of $\be$ can be understood intuitively as follows. Since the $\be$-term in the Hamiltonian can be rewritten through $\sum_i k_i^2$, where $k_i\equiv \sum_j c_{ij}$ is the degree of vertex $i$, a large negative $\be$ strongly suppresses high vertex degrees. Then, since the mean degree is kept at the desired value by adjusting $\al$, it implies that degrees below the mean value are also strongly suppressed, leading to a small degree variance. Detailed analysis can be found in \cite{AC}.

The two-star model by itself does not possess any significant geometric properties (for instance, its triangle count is very low). We shall then use it as a point of departure for further modifications and look for Hamiltonians of the form
\beq
H(c)=H_{2*}(c)-\s H_\mathrm{mod}(c),
\label{Hamgen}
\eeq
trying to find $H_\mathrm{mod}$ that generates an ensemble with the desired properties. One could naively choose $H_\mathrm{mod}$ as the total number of triangles in the graph, given by $\Tr(c^3)/6$, but this is well-known to fail since the early days of random graph simulations \cite{strauss, PN2}.
If one tries this guess with $\beta=0$, the triangle term, whose maximal value goes like $O(N^3)$ is capable of overpowering in the thermodynamic limit the $\al$-term which goes at most as $O(N)$. As a result, the model forms two phases, one very dense and one very sparse, and it is impossible to tune the parameters to attain
reasonable vertex degrees and high numbers of triangles at the same time. If one tries to use a large negative $\beta$, as described above, to make the graph approximately regular, it is possible to attain large numbers of triangles in such an approximately regular graph, but at the cost of another terminal pathology: the graph bursts into a collection of disconnected nearly complete graphs with just a few vertices each. (Bursting into similar, somewhat more rarefied disconnected components has been referred as creation of `baby universes' in the context of \cite{KT,KTB} following the spirit of the simplicial quantum gravity literature \cite{QG,CDT}.) 
\begin{figure}[t]
	\centering
	\includegraphics[width = \textwidth]{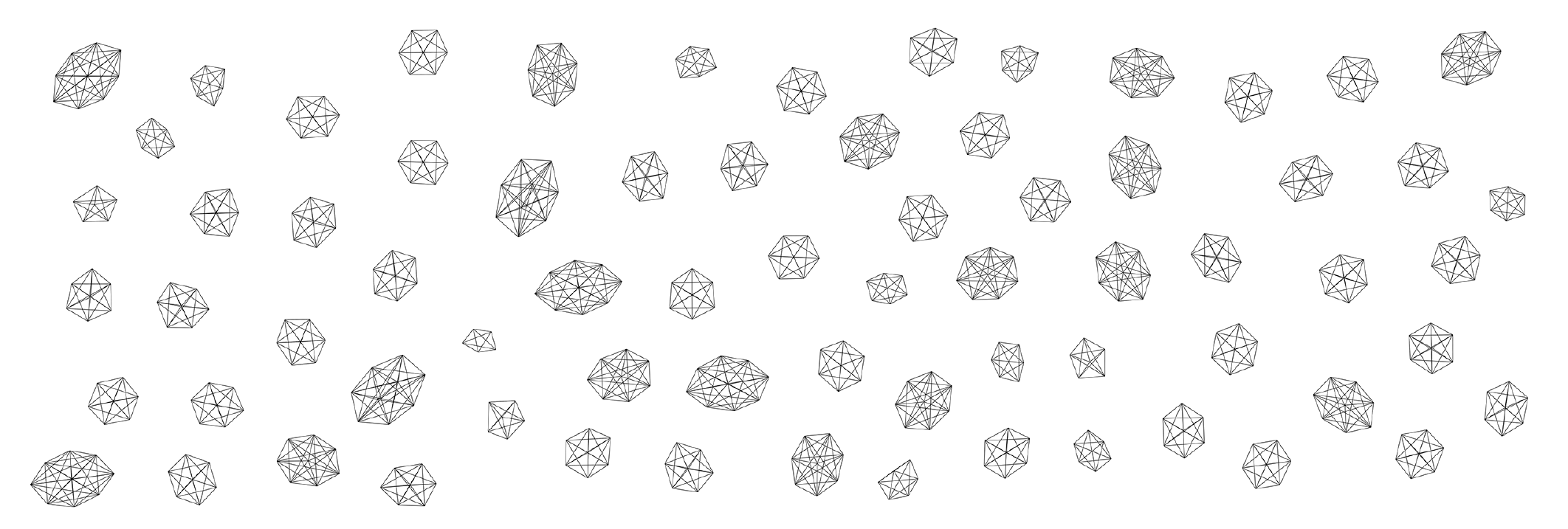}
	\caption{A fractured phase graph with $N=500$ vertices obtained from simulating (\ref{Hamgen}) where a large negative $\beta$ is used to keep the degree variance small and $H_\mathrm{mod}$ is chosen to be the number of triangles. The specific values of the control parameters are $\al=170$, $\be=-20$, $\s=10$. There is a stark contrast between this fractured graph and the fully connected graphs with large numbers of triangles produced by the new models we develop below.}
	\label{dust}
\end{figure}
Such dust-like configurations (an example is given in Fig.~\ref{dust}) are of little interest insofar as one is concerned with approaching large-volume discretized spaces. One can gain some intuitive understanding of the above fracturing phenomenon by counting triangles in $d$-regular graphs. The maximal number of triangles formed around a given vertex in such a graph is $d(d-1)/2$ (assuming that every pair of edges emerging from this vertex is completed into a full triangle). For a graph with N vertices, since each triangle is shared among 3 vertices, this translates into $Nd(d-1)/6$ triangles. But this bound is saturated by splitting the graph into $N/(d+1)$ disconnected complete graphs with $(d+1)$ vertices, and hence $d(d+1)(d-1)/6$ triangles, each.

Models with the triangle count in $H_\mathrm{mod}$ replaced by the counts of other similar motifs (for example, squares) meet a similar fate.
They tend to suffer condensation transitions if the vertex degree is unconstrained, and fracturing if the degree is constrained (either rigidly, or by a large negative $\be$-term in the graph Hamiltonian). As an aside, counting formulas for arbitrary polygons (cycles of arbitrary length) in terms of the adjacency matrix have been systematically explored in the literature \cite{HM,mw}.

One thus has to be more ingenious and use modifying terms in (\ref{Hamgen}) that do not simply introduce a Lagrange
multiplier for the number of specific motifs, and yet succeed with generating
graph ensembles that enjoy large numbers of such motifs. Some models of this form have been proposed
in \cite{newexp}, but they rely on rather complicated graph Hamiltonians and do not aim at producing
graphs of approximately constant degree. While the models we shall formulate show some affinity to
the models of \cite{newexp}, our main inspiration comes from the observation in \cite{KT} that
one can prevent condensation and fracturing in a graph model with a simple square-counting Hamiltonian
if one forces the graphs to be bipartite and regular, and satisfy the {\it hard-core condition}. This hard-core condition is crucial in the construction and says that no two squares (cycles of length 4) may possess more than one edge in common. In particular, the hard-core condition allows subgraphs of the form
\SetGraphUnit{1}
\GraphInit[vstyle=Simple]
\SetVertexSimple[MinSize    = 8pt]
\beq
\begin{split}
\begin{tikzpicture}
  \GraphInit[vstyle=Simple]
  \Vertices{circle}{A,B,C,D}
  \Edges(A,B,C,D,A)
\end{tikzpicture},
\end{split}
\label{1sq}
\eeq
but forbid subgraphs of the form\
\beq
\begin{split}
\begin{tikzpicture}
  \GraphInit[vstyle=Simple]
  \Vertices{circle}{A,B,C,D}
  \Vertex[x=0.3,y=0]{E}
  \Edges(A,B,C,D,A)
  \Edges(B,E,D)
\end{tikzpicture}.
\end{split}
\label{3sq}
\eeq
We would like to avoid introducing such hard constraints (and keep nonzero, possibly small, probabilities for all conceivable graphs), but we can mimic this constraint in the context of the exponential random graph model (\ref{Hamgen}) by choosing $H_\mathrm{mod}$ that favors (\ref{1sq}) but not (\ref{3sq}). (Note that, in a simple graph, two squares may share 0, 1 or 2 edges. Hence, encouraging the production of simple squares (\ref{1sq}) but not (\ref{3sq}) effectively encourages production of squares that share exactly one edge.) Thus, for every pair of vertices, we shall give a reward if there are exactly two paths of length 2 between them. Formally, this is most conveniently expressed by first introducing the square of the adjacency matrix
\beq
q_{ij}=\sum_k c_{ik}c_{kj},
\label{csq}
\eeq
which simply counts the number of paths of length 2 between vertices $i$ and $j$. We then write
\beq
H_\mathrm{mod}= H_\mathrm{sq}\equiv\sum_{i>j} \de(q_{ij},2),
\label{Hsq}
\eeq
where $\de$ is the ordinary Kronecker symbol:
\beq
\de(n,m)=1\,\,\,\mathrm{ if }\,\,\, n=m,\,\,\, \mathrm{ and }\,\,\, 0\,\,\,\mathrm{ otherwise}.
\label{Kronecker}
\eeq
Mnemonically, (\ref{Hsq}) is simply the number of entries 2 in the (off-diagonal) upper triangle of the matrix square of the adjacency matrix $c$.
Configurations of the form (\ref{1sq}-\ref{3sq}) belong to the class called $k$-independent two-paths in \cite{newexp}, though the way they are used in our action is different from \cite{newexp}.

The model defined by (\ref{Hsq}) is adapted to producing squares while avoiding clumping/frac\-turing, and we shall see in the next section that it is precisely what it does. Note, in particular, that the small complete $d$-regular graphs that underlie fracturing as explained in the passage under (\ref{Hamgen}) are not favored by (\ref{Hsq}), while they would have been favored by a naive $H_\mathrm{mod}$ counting the number of squares.

Furthermore, one can straightforwardly modify (\ref{Hsq}) to make it adapted to production of triangles rather than squares. Namely, we want a Hamiltonian that favors subgraphs of the form
\beq
\begin{split}
\begin{tikzpicture}
  \GraphInit[vstyle=Simple]
  \Vertices{circle}{A,B,C,D}
  \Edges(A,B,C,D,A)
  \Edges(B,D)
\end{tikzpicture},
\end{split}
\label{2tr}
\eeq
which contain two triangles, but does not favor subgraphs of the form
\beq
\begin{split}
\begin{tikzpicture}
  \GraphInit[vstyle=Simple]
  \Vertices{circle}{A,B,C,D}
  \Vertex[x=0.45,y=0]{E}
  \Edges(A,B,C,D,A)
  \Edges(B,E,D,B)
\end{tikzpicture},
\end{split}
\label{3tr}
\eeq
which would have encouraged fracturing/clumping (more than two triangles share an edge). This is accomplished by choosing
\beq
H_\mathrm{mod}= H_\mathrm{tr}\equiv\sum_{i>j} \de(c_{ij}q_{ij},2),
\label{Htr}
\eeq
This is just the number of entries 2 in the upper triangle of the Hadamard (entrywise) product of the adjacency matrix $c$ and its ordinary matrix square $q$. Configurations of the form (\ref{2tr}-\ref{3tr}) belong to the class called $k$-triangles in \cite{newexp}, though the way they are used in our action is different from \cite{newexp}.


\section{Numerical simulations}\label{sec: numerical}

We now set out to explore the models defined by (\ref{expect}-\ref{Hamgen}) and (\ref{Hsq}) or (\ref{Htr}).
While the graph Hamiltonians are rather simple, and at some level are only slightly more complex than the much-studied two-star model,
the most viable approach is to analyze the corresponding graph ensembles using computer simulations.
We shall give some preliminary comments on the analytic structure of the models in the next section.

\subsection{Remarks on the Monte Carlo implementation}

General discussion of random graph simulations using Monte Carlo techniques can be found in \cite{CAR,MCMC}. We use a simple sequence of stochastic flips that randomly select a pair of distinct vertices
$i$ and $j$, and then either erase the edge between them, if it exist, or add an edge, if there is none to start with.
The change of the adjacency matrix under such an $(ij)$-flip can be written as
\beq
\Delta_{ij} c_{mn}= (\de_{im}\de_{jn}+\de_{in}\de_{jm})(1-2c_{ij}).
\eeq
As usual in Metropolis-type algorithms, this proposed $(ij)$-flip is accepted with probability 
$\min(1, e^{H(c)-H(c+\Delta_{ij}c)})$.

For an efficient implementation of this algorithm, it is important to optimize the computation of the
variation of the graph Hamiltonian (\ref{Hamgen}) under the flip. The two-star part (\ref{2st}) is very
simple, as it can be completely expressed through the vertex degree sequence
\beq
k_n=\sum_m c_{mn}.
\eeq
We can now rewrite (\ref{2st}) as
\beq
H_{2*} = -\al \sum_{i} k_i - \be \sum_{i} k_i^2.
\eeq
The variation of the degree sequence is simply
\beq
\Delta_{ij} k_n=(\de_{in}+\de_{jn})(1-2c_{ij}).
\label {var_k}
\eeq
It then follows \cite{CAR} from (\ref{var_k}) that the variation of the two-star Hamiltonian is
\beq
\Delta_{ij} H_{2*} = -2(1-2c_{ij})(\al + \be (k_i + k_j)) + 2\be.
\eeq

What remains is to explain how to compute the variation of $H_\mathrm{mod}$. We start with the case of the `square' model (\ref{Hsq}), which is slightly simpler. Since $H_\mathrm{sq}$ is expressed through $q$, the square of the adjacency matrix, it suffices to write down the variation of $q$ under the flip:
\beq
\Delta_{ij} q_{mn}= (\de_{im}c_{jn}+\de_{jm}c_{in}+\de_{jn}c_{im}+\de_{in}c_{jm})(1-2c_{ij})
+(\de_{im}\de_{in}+\de_{jm}\de_{jn}).
\label{delqmn}
\eeq
All the changes in $q$ thus occur in rows and columns number $i$ and $j$. Without recomputing the whole matrix, which would have been costly for large $N$, one can simply recompute these two rows. Comparing the number of entries 2 in the old rows and the new ones gives the variation $H_\mathrm{sq}$, and,
together with the variation of $H_{2*}$, can be converted into the flip acceptance probability. If the flip is accepted, the corresponding rows in $q$, which is stored as an explicit matrix, are updated. In this way,
the number of operations required per flip grows very moderately as $O(N)$. One could in fact try to use the sparsity of the adjacency matrix and its square in the low density regime that we focus on to further reduce the number of operations per step to $O(1)$, but this would require dealing with sparse matrix formats that have their own disadvantages, and we shall avoid it.

A similar prescription can be given for the modified triangle model (\ref{Htr}). The Hamiltonian is expressed
through the  products $c_{mn}q_{mn}$ whose variation is given by
\beq
\Delta_{ij}(c_{mn} q_{mn}) = q_{mn} (\Delta_{ij} c_{mn}) + c_{mn} (\Delta_{ij} q_{mn}) + (\Delta_{ij} c_{mn})(\Delta_{ij} q_{mn}).
\label{delcmnqmn}
\eeq
Note that $\Delta_{ij} c_{mn}$ is only non-zero if $m=i$ and $n=j$ or $m=j$ and $n=i$, in which case $\Delta_{ij} q_{nm}$ is zero, hence the last term drops out. The term $c_{mn}(\de_{im}\de_{in}+\de_{jm}\de_{jn})$ from the expansion of the second term in (\ref{delcmnqmn}), upon the substitution of (\ref{delqmn}), also vanishes because the diagonal elements of $c$ equal zero. Thus, (\ref{delcmnqmn}) can be simplified to
\beq
\Delta_{ij}(c_{mn} q_{mn}) = q_{mn} (\de_{im}\de_{jn}+\de_{in}\de_{jm})(1-2c_{ij}) + c_{mn}  (\de_{im}c_{jn}+\de_{jm}c_{in}+\de_{jn}c_{im}+\de_{in}c_{jm})(1-2c_{ij}).
\eeq
As before, all the changes of the matrix $\tilde q$ defined by $\tilde q_{mn}=c_{mn}q_{mn}$ are in rows and columns number $i$ and $j$. The way to proceed then is to store in memory the matrices $c$, $q$ and $\tilde q$, updating after each flip the relevant rows and columns, which requires $O(N)$ operations per step.

As laid out among our motivations, we intend to obtain graphs of roughly constant small degree. We choose this reference degree to be 6, which for the triangle model is expected to encourage formation of pieces of two-dimensional triangular lattice, and for the square model, pieces of three-dimensional cubic lattice. The degree variance is kept low by taking a large negative value of the two-star parameter $\beta$. One then chooses $\sigma$ as one wishes, but $\alpha$, the Lagrange multiplier for the number of edges, has to be adjusted to make the mean degree close to 6. This is in practice implemented by running a sort of `adaptive' Monte Carlo simulation, in which, every now and then, the current mean degree is measured and $\alpha$ is adjusted in proportion to how much it deviates from 6. Once this `adaptive' simulation has converged, one can take the resulting value of $\alpha$ and rerun a completely conventional Monte Carlo simulation with $\alpha$ fixed at that value. All the numerical results we report below are obtained in this manner by running
conventional Monte Carlo simulations after adaptive Monte Carlo simulations have been used for initial estimation of the desired values of the thermodynamic parameters.

It is a separate question how the parameters $\alpha$, $\beta$, $\sigma$ should depend on $N$ for comparing the simulations at different $N$. Our numerical experiments indicate that keeping $\beta$ and $\sigma$ fixed and choosing the value of $\alpha$ that brings the mean degree close to 6 results in approximately the same number of geometric features, to be defined below, in proportion to $N$. One expects that the value of $\alpha$ necessary for maintaining a small fixed degree varies logarithmically with $N$, as in the two-star model \cite{AC}, but this variation is rather small in any case for the range of $N$ we consider (200 to 5000), and we do not keep track of it explicitly. (The value of $\alpha$ is thus simply chosen at each $N$ for the given $N$-independent $\beta$ and $\sigma$ to ascertain that the mean degree is close to 6.)

\subsection{The modified triangle model}

We have performed simulations of the modified triangle model defined by (\ref{2st}), (\ref{Hamgen}) and (\ref{Htr}), and observed that it stands up to our expectations. In an appropriate parameter regime that we shall specify shortly, the model forms a very large number of triangles and bigger geometric structures, without experiencing any clumping or fracturing that plague naive Strauss-type models.

We first state our general approach to quantifying the amount of geometric structure in the graphs.
To enable comparisons between graphs of different sizes, it is convenient to introduce ratios of
the observed number of geometric primitives to some reference numbers for a graph of a given size.
For these reference numbers, we choose estimates for the maximal number of times a particular primitive can be found in a graph with $N$ vertices. In some cases, as with the hexagon count defined precisely below, this maximal number is known with certainty. Where we do not have a rigorous upper bound, we use the values for the fractured 6-regular graph consisting of $N/7$ complete 6-regular graphs with 7 vertices each. This graph, for example, provides a rigorous upper bound on the number of triangles in a 6-regular graph, as explained previously, and while our graphs are not exactly 6-regular, their degree variance is small, and the estimate based on 6-regular graphs is expected to give a good sense of the maximal possible number of triangles.

One might wonder why we are not quantifying the number of geometric primitives by comparing the observed values in our model to another simpler random graph model. The reason is that simple models like the Erd\H{o}s-R\'enyi or two-star do not produce any noticeable numbers of geometric primitives at all (the fractions we define below are all zero for these models in the $N\to\infty$ limit at finite mean degree). Other known models like the Strauss triangle model do produce large numbers of geometric primitives but only in the fractured phase that is of no interest from our geometric perspective, and that undermines meaningful comparisons with those models. The best solution we find is to benchmark our model in relation to estimates for the maximal number of geometric primitives of a given sort in a random graph with the relevant degree range.

Following the above guidelines, we introduce the triangle fraction $\eta_\Delta$, given by the number of triangles \cite{HM,mw} $n_\Delta=\Tr(c^3)/6$, divided by the maximal number of triangles in a $d$-regular graph, given by $Nd(d-1)/6$. Since we tune $\alpha$ and $\beta$ to make our graphs approximately 6-regular, $Nd(d-1)/6$ effectively equals $5N$, and hence
\beq
\eta_\Delta=\frac{n_\Delta}{5N}.
\label{etatri}
\eeq
We additionally count the number of hexagons, defined as subgraphs of the form
$$
\begin{tikzpicture}
  \GraphInit[vstyle=Simple]
  \Vertices{circle}{A,B,C,D,E,F}
  \Vertex[x=0,y=0]{G}
  \Edges(A,B,C,D,E,F,A)
  \Edges(A,G,B)
  \Edges(C,G,D)
  \Edges(F,G,E)
\end{tikzpicture},
$$
with the extra requirement that the degree of the central vertex is precisely six (which means it has no connections to vertices outside this subgraph), and the peripheral vertices do not have any extra connections to each other besides the one displayed (while their connections to the rest of the graph are unrestricted). Since each graph vertex may lie at the center of at most one hexagon as defined, it is natural to introduce the hexagon fraction relative to the number of vertices in terms of the hexagon number $n_{hex}$ as
\beq
\eta_{hex}=\frac{n_{hex}}N.
\label{etahex}
\eeq
Our choice to count hexagons, as opposed to other triangle-based motifs, is suggested by the expectation that, with the dominant degree 6 and a Hamiltonian that discourages more than two triangles to share an edge, pieces of two-dimensional hexagonal lattice will form, which is indeed what we observe.

\begin{figure}[t]
    \noindent
    \centering
\begin{tikzpicture}
    \node(img){\includegraphics[width = 0.95\textwidth]{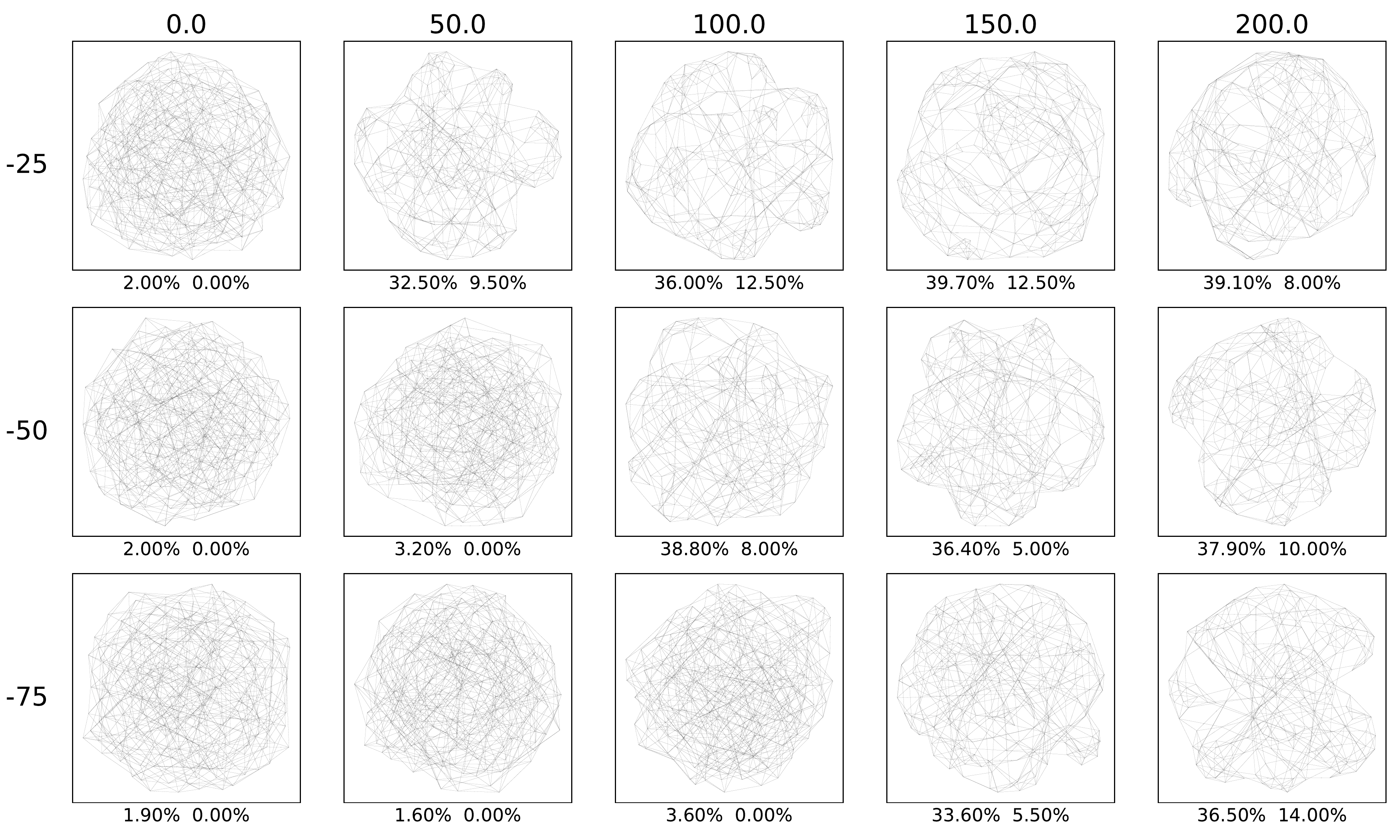}};
    \node[above=of img, node distance=0cm, yshift=-1cm] {$\sigma$};
    \node[left=of img, node distance=0cm,anchor=center, xshift=0.7cm] {$\beta$};/
\end{tikzpicture}
    \caption{Random graphs of the modified triangle model at $N=200$ for a range of values of $\beta$ and $\sigma$, with $\alpha$ adjusted so that the mean degree is close to 6 (the actual values are within the range $6\pm0.1$). The percentages under each graph represent the triangle fraction (\ref{etatri}) and the hexagon fraction (\ref{etahex}), respectively. We remind the readers that increasing the magnitude of $\beta$ (moving down the vertical axis) serves to decrease the degree variance, while increasing $\s$ (moving to the right horizontally) further encourages the production of triangles.}
    \label{fig:tri_scan}
\end{figure}

With these preliminaries, we now set out to explore the parameter space of our model. A quick impression can be obtained from Fig.~\ref{fig:tri_scan}, 
where we have chosen a relatively small number of vertices ($N=200$) to make the visualization transparent in a regular printout, though the picture is qualitatively similar for the higher values of $N$ that we have explored (up to 5000). 
What one notices immediately in visual terms is the presence of two distinct regions. If $\sigma$ is insufficiently large (below  the diagonal), the graph is visually threadball-like, with very small triangle counts and no hexagons. Above the diagonal, on the other hand, the graphs have a crystalline appearance with a large number of triangles and appreciable number of hexagons.\footnote{Graphs in this region visually appear rather similar to some of the graphs in \cite{W2}, which are deterministically generated. This is perhaps simply a reflection of the fact that triangle relations play a role in the formulations both here and in \cite{W2}.} We emphasize that the observed number of triangles (a third, or more, of the maximal number possible at this vertex degree) is extremely large, expressing a high `transitivity' in the language of Strauss \cite{strauss}. By contrast, in the simplest Erd\H{o}s-R\'enyi graph with a fixed small ($N$-independent) mean vertex degree, the fraction of triangles tends to 0 for large graphs, while the probability of finding even a single hexagon is 0. For the rest of our treatment, we shall focus on graphs with strong geometric features, above the diagonal in Fig.~\ref{fig:tri_scan}. Note that it would be interesting to explore the phase transition that presumably separates the two regions in Fig.~\ref{fig:tri_scan}, but it is a topic apart from our main objectives here.

\begin{figure}[t]
	\centering
	\includegraphics[width = 0.7\textwidth]{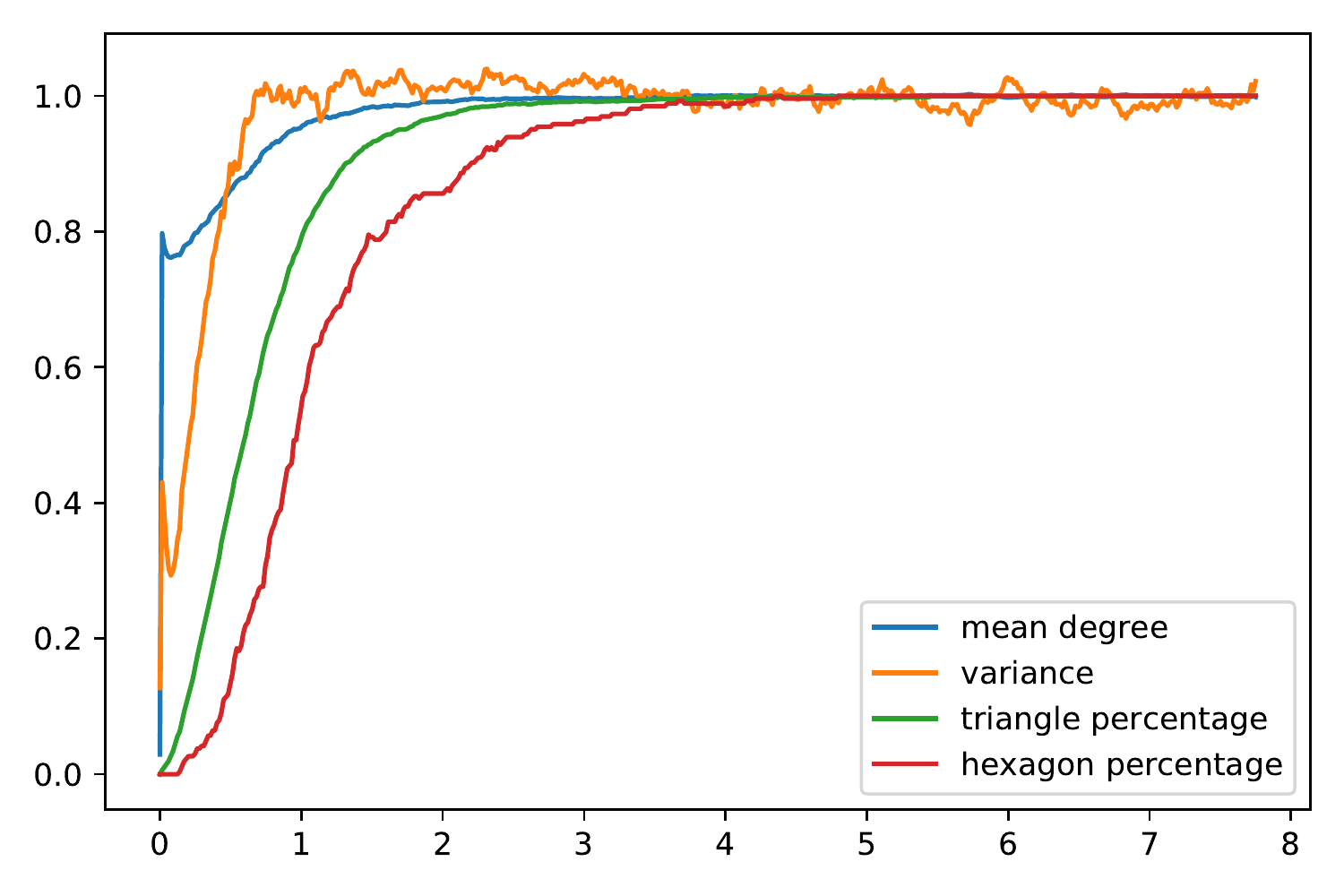}
	\caption{Convergence of the mean degree, degree variance, and the fractions of triangles and hexagons at $N=2000, \alpha=177, \beta=-20, \sigma=100$, starting from an empty graph initial state. Each curve is rescaled so that the final steady-state value is one. The horizontal axis counts the elementary Monte Carlo steps in units of 10 million.}
	\label{fig:con_plot}
\end{figure}

We comment on the convergence of our Monte Carlo simulations, exemplified by Fig.~\ref{fig:con_plot}, now for a larger graph with $N=2000$. 
One observes that the mean degree converges first, followed by the degree variance, while the geometric structures form and stabilize later. (Such multiple equilibration scales have been seen in a related context in \cite{NRS}.) The mean degree and degree variance are defined as $\bar k\equiv\sum_i k_i/N$ and $\mbox{v}_k\equiv\sum_i k_i^2/N - (\sum k_i)^2/N^2$, while the triangle and hexagon fractions are defined as in (\ref{etatri}) and (\ref{etahex}). We have performed simulations at various values of $N$ and our rough estimate is that the number of Monte Carlo steps necessary for the number of hexagons to stabilize grows like  $O(N^{2.8})$.

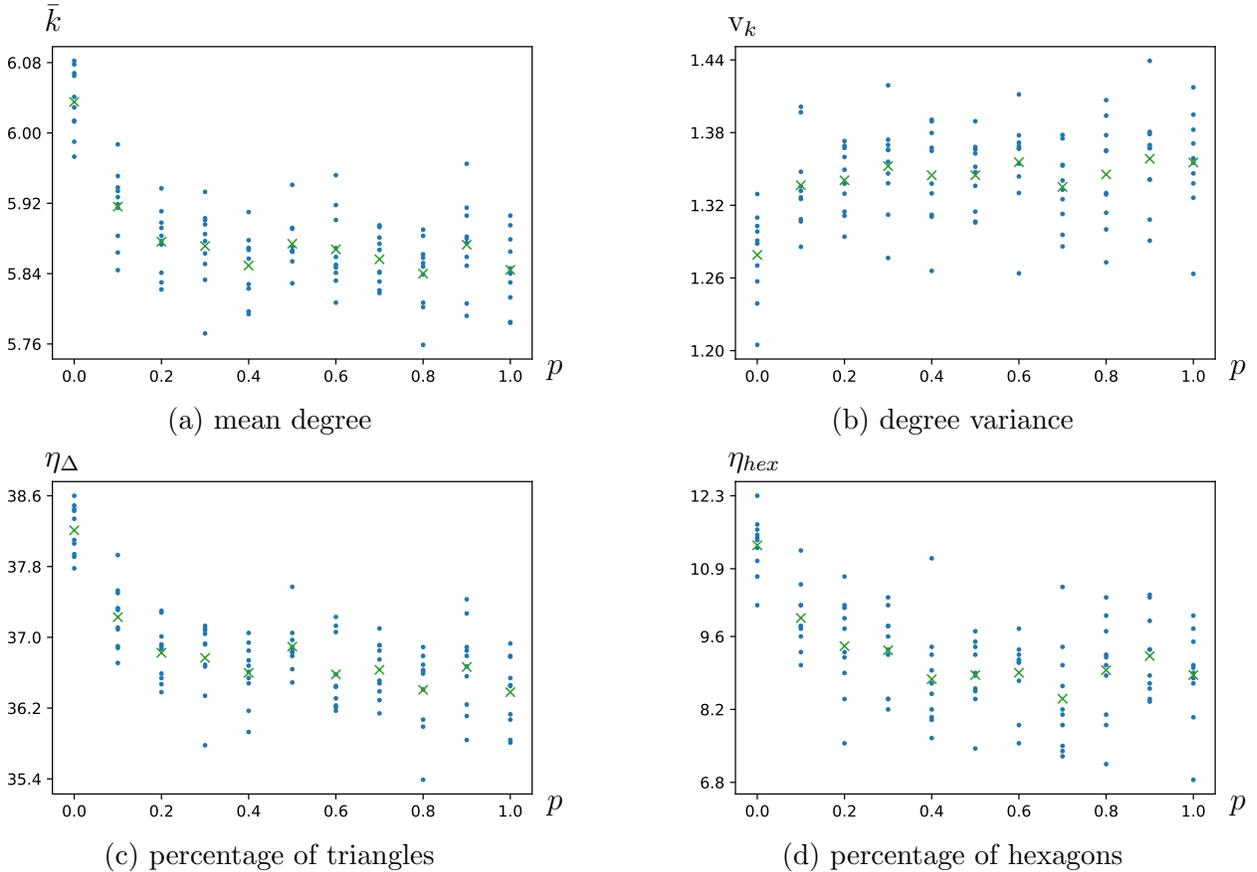
\begin{figure}[t]
    \centering
    \begin{subfigure}[b]{0.45\textwidth}
         \centering
          \includegraphics[width =\textwidth]{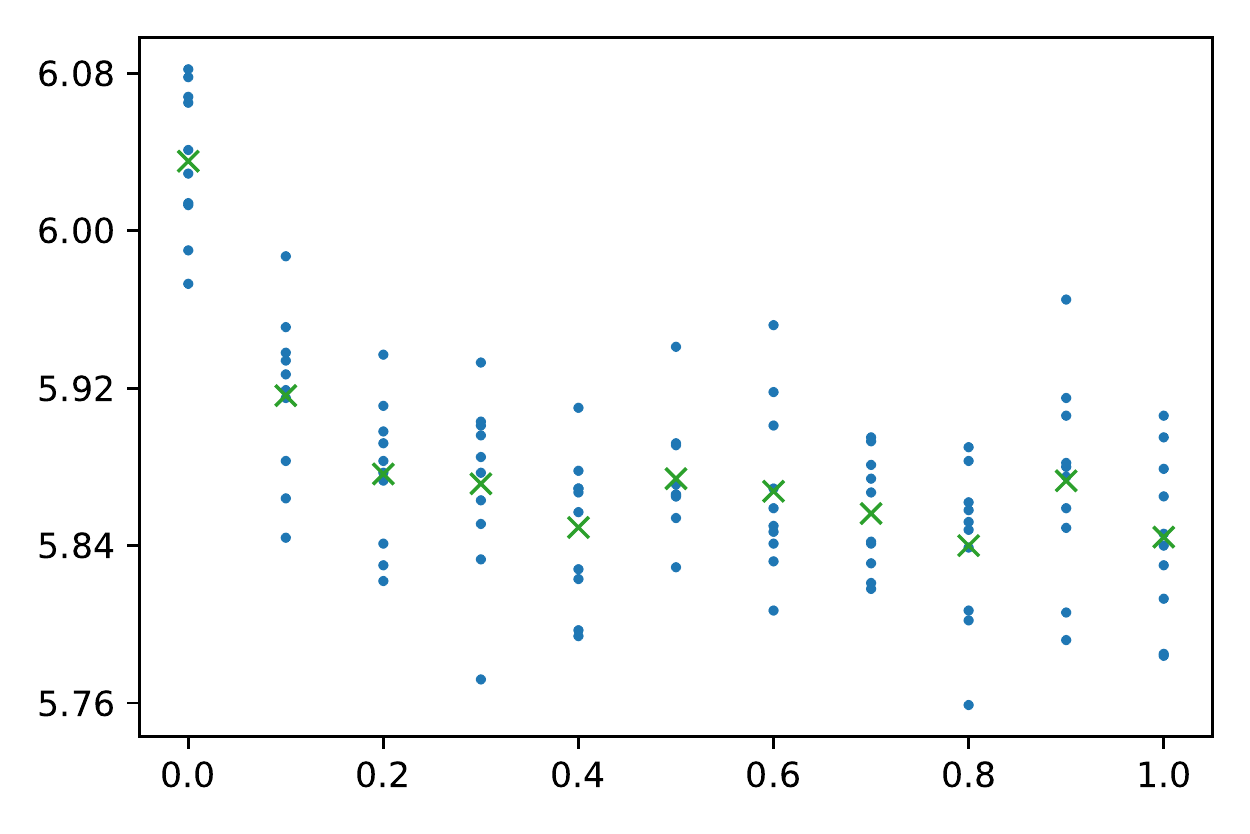}
\begin{picture}(0,0)
\put(-85,155){$\bar k$}
\put(105,25){$p$}
\end{picture}\vspace{-7mm}
         \caption{mean degree\vspace{3mm}}
         \label{fig:tri_deg}
    \end{subfigure}
    \hfill
    \begin{subfigure}[b]{0.45\textwidth}
         \centering
          \includegraphics[width =\textwidth]{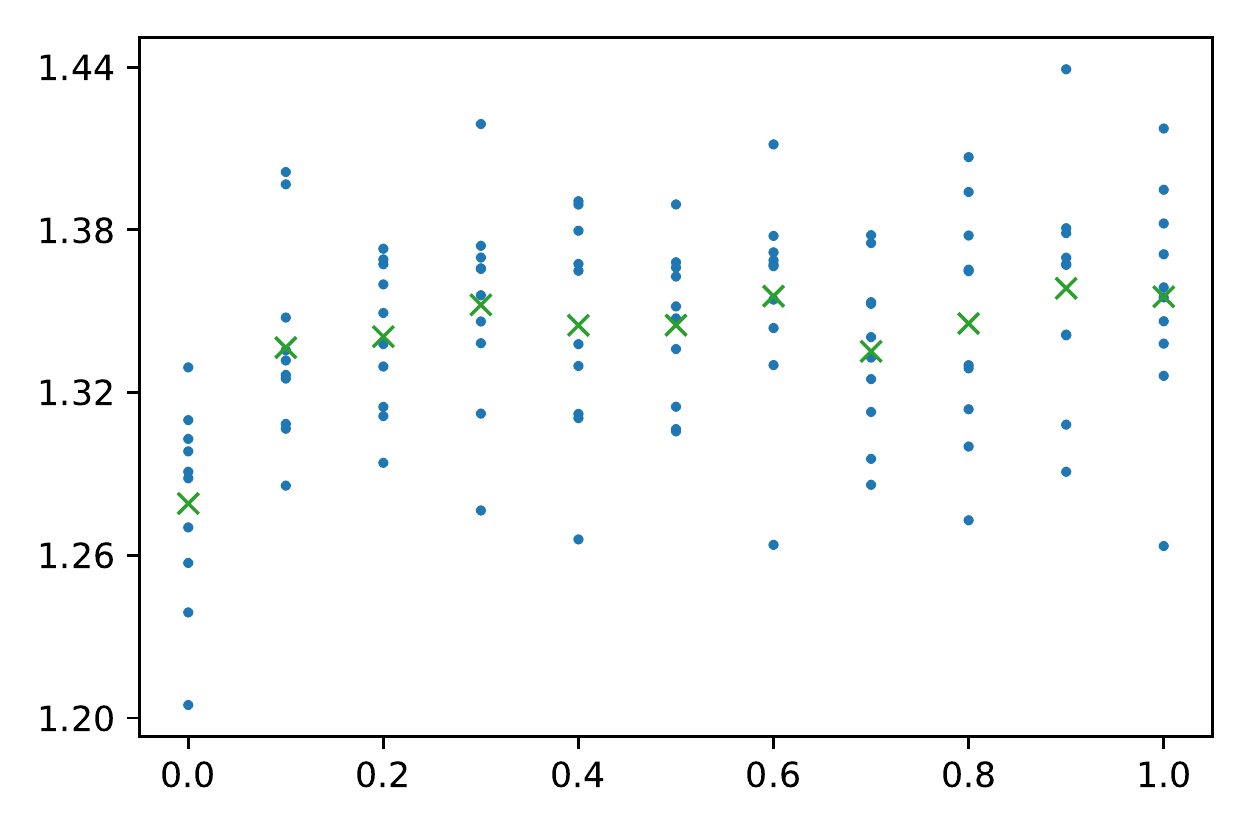}
\begin{picture}(0,0)
\put(-85,155){$\mbox{v}_k$}
\put(105,25){$p$}
\end{picture}\vspace{-7mm}
         \caption{degree variance\vspace{3mm}}
         \label{fig:tri_var}
    \end{subfigure} 
    \\
    \begin{subfigure}[b]{0.45\textwidth}
         \centering
          \includegraphics[width =\textwidth]{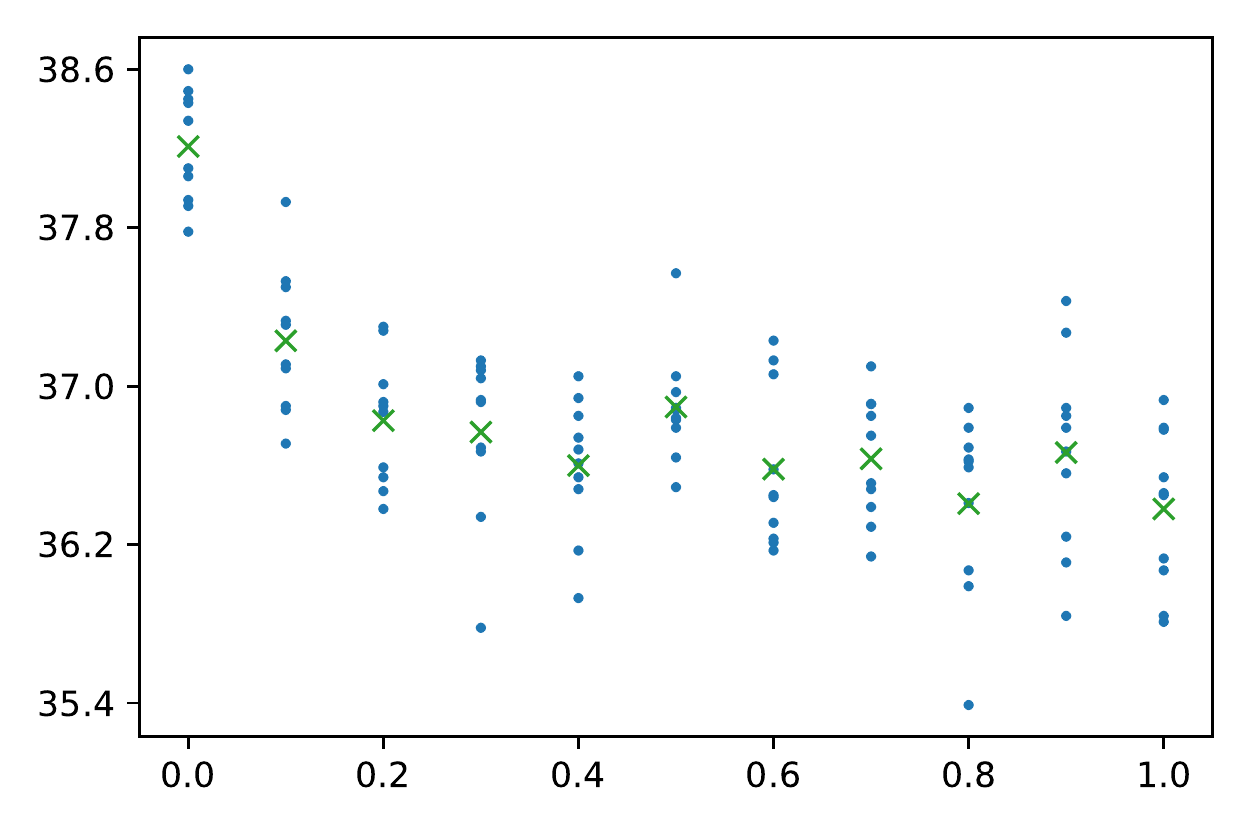}
\begin{picture}(0,0)
\put(-85,155){$\eta_\Delta$}
\put(105,25){$p$}
\end{picture}\vspace{-7mm}
         \caption{percentage of triangles\vspace{2mm}}
         \label{fig:tri_sqr}
    \end{subfigure} 
    \hfill
    \begin{subfigure}[b]{0.45\textwidth}
         \centering
          \includegraphics[width =\textwidth]{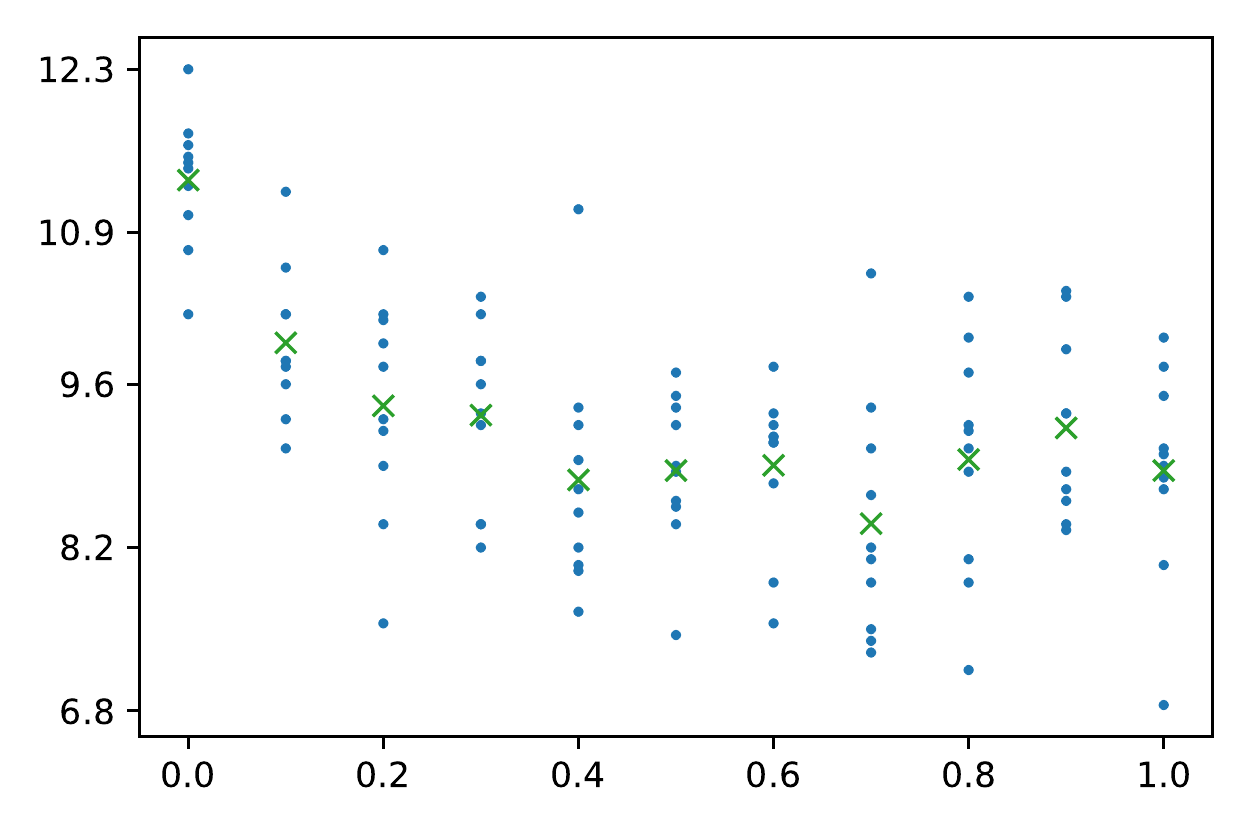}
\begin{picture}(0,0)
\put(-85,155){$\eta_{hex}$}
\put(105,25){$p$}
\end{picture}\vspace{-7mm}
         \caption{percentage of hexagons\vspace{2mm}}
         \label{fig:tri_cube}
    \end{subfigure}
    \caption{Outcomes of multiple simulations of the modified triangle model at $N=2000, \alpha=177, \beta=-20, \sigma=100$, with initial states in the form of Erd\H{o}s-R\'enyi graphs whose edge occupation probability $p$ is given on the horizontal axes. The green crosses indicate the mean of each vertical column.}
    \label{fig:tri_plots}
\end{figure}

In the regime that interests us, namely, a large value of $\sigma$ and a developed geometric structure in the graphs, the model displays jamming behaviors, and the values of observables at which the Monte Carlo evolution stabilizes vary somewhat between different runs with the same values of thermodynamic parameters. This is perhaps not very surprising since, if the graphs have a `crystalline' appearance, they form in a process of spontaneous crystallization, where jammed behaviors often occur. Indeed, once a graph has formed with a large number of configurations of the form (\ref{2tr}), which our Hamiltonian favors, it becomes difficult to move edges around without upsetting this order, so as to achieve even more optimal configurations. In practice, as seen from Fig.~\ref{fig:con_plot}, the number of hexagons, for example, simply freezes after a sufficiently long run, and stops fluctuating altogether on any time scales our
numerical simulations can reasonably access. In order to quantify the variations in these final jammed configurations and the failure to reach the true equilibrium, we have prepared Fig.~\ref{fig:tri_plots}
which summarizes the outcomes of multiple runs with the same thermodynamic parameters, but starting with initial configurations given by Erd\H{o}s-R\'enyi graphs of different densities. (For each given initial density, we restart the simulations many times, each time initializing the graph with a newly constructed Erd\H{o}s-R\'enyi graph having that edge density.) One can see that the mean degree converges to a predictable value within a narrow range, while the degree variance, and the triangle (\ref{etatri}) and hexagon (\ref{etahex}) fractions display greater variation, with the hexagon fraction being particularly variable.
Irrespectively of that, all of the graphs that come out of our simulations contain geometric primitives (triangles and hexagons in our case) at rates orders of magnitude higher than what one could hope to obtain
from exponential random graph models without the extra terms we have designed, which is what we see as a quantification of the success of our model.

\begin{figure}[t!]
	\centering\vspace{-5cm}
	\includegraphics[width = \linewidth]{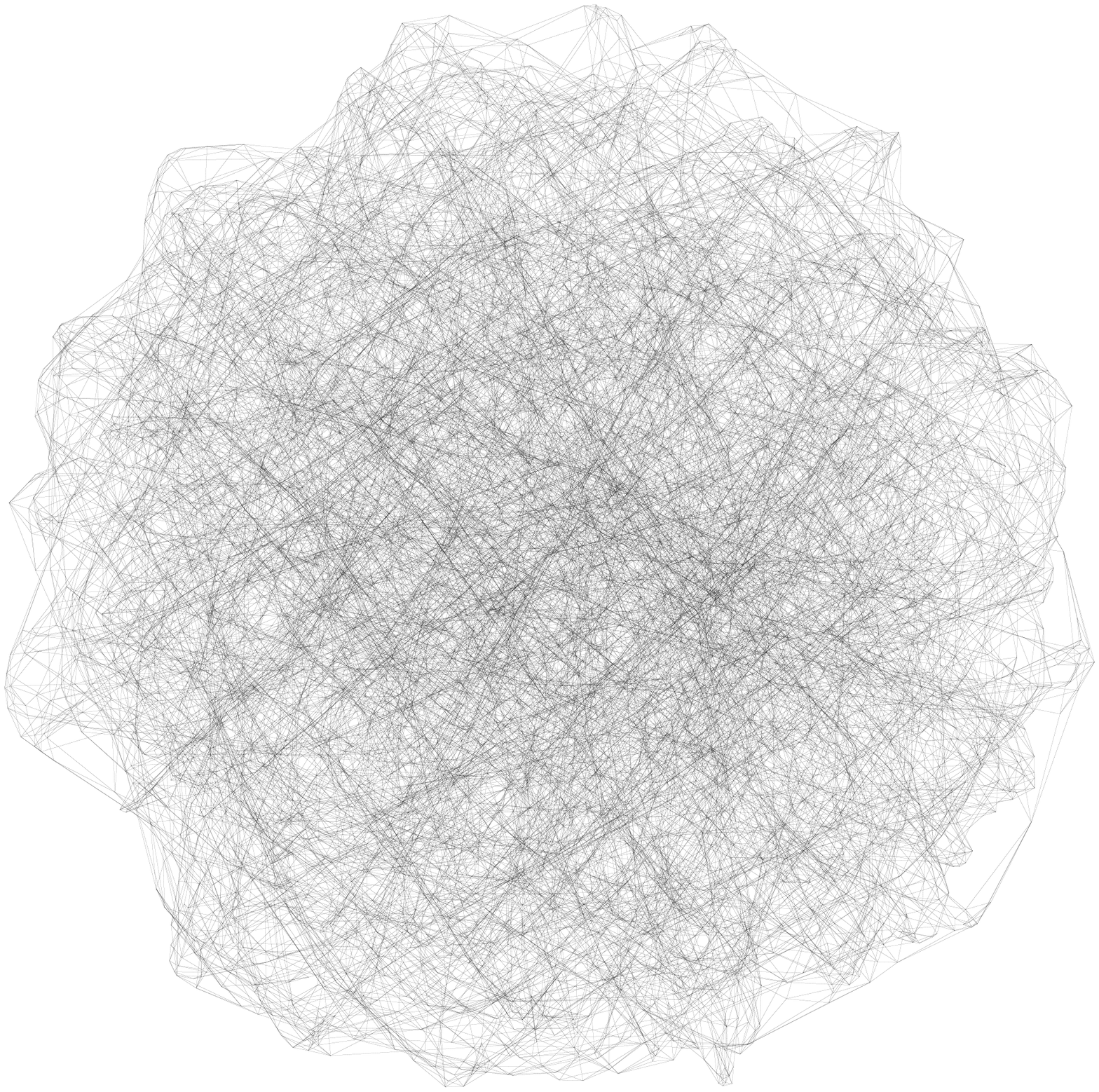}\vspace{-3cm}
	\caption{A large graph obtained from simulations of the modified triangle model at $N=5000, \alpha=177, \beta=-20, \sigma=100$. The triangle and hexagon fractions are very high, at around 38\% and 11\% respectively. Due to the high density of such a large graph in a regular page presentation, the geometric structure is less evident to the naked eye than for the similar graphs with a much smaller number of vertices in the upper right part of Fig.~\ref{fig:tri_scan}, but it is more discernible at the periphery.}
	\label{fig:big_tri}
\end{figure}

To conclude, we present in Fig.~\ref{fig:big_tri} a particular very large graph with $N=5000, \alpha=177, \beta=-20, \sigma=100$ that has come out of our simulations, in order to give the readers an impression of its visual qualities.

\subsection{The modified square model}

The modified square model defined by (\ref{2st}), (\ref{Hamgen}) and (\ref{Hsq})
displays features very similar to what we have described above for the triangle model,
with the role of triangles taken up by squares, and the role of hexagons, by cubes.

The number of squares can be computed as \cite{HM,mw} 
$n_{\Box}=(\Tr(c^4)+\Tr(c^2)-2\sum_i k_i^2)/8$.
We choose to normalize this number by the number of squares in a disconnected collection of $N/(d+1)$ complete $d$-regular graphs with $(d+1)$ points each, given by $Nd(d-1)(d-2)/8$, which becomes $15N$ for $d=6$, yielding our definition
\beq
\eta_{\Box}=\frac{n_\Box}{15N}.
\eeq
We furthermore consider the number of cubes defined as subgraphs of the form
$$
\begin{tikzpicture}
  \GraphInit[vstyle=Simple]
  \Vertex[x=0,y=0]{A}
  \Vertex[x=0,y=1]{B}
  \Vertex[x=1,y=1]{C}
  \Vertex[x=1,y=0]{D}
  \Vertex[x=0.4,y=0.4]{E}
  \Vertex[x=0.4,y=1.4]{F}
  \Vertex[x=1.4,y=1.4]{G}
  \Vertex[x=1.4,y=0.4]{H}
  \Edges(A,B,C,D,A)
  \Edges(E,F,G,H,E)
  \Edges(A,E)
  \Edges(B,F)
  \Edges(C,G)
  \Edges(D,H)
\end{tikzpicture}\,\,\,,
$$
in which the eight depicted vertices are connected to each other exactly as given, without any extra connections, but they may be connected to the rest of the graph in an arbitrary way. We then compute the number of cubes $n_{cube}$ and define the cube fraction in proportion to the total number of vertices
\beq
\eta_{cube}=\frac{n_{cube}}N.
\eeq
\begin{figure}[t]
    \centering
    \begin{subfigure}[b]{0.45\textwidth}
         \centering
          \includegraphics[width =\textwidth]{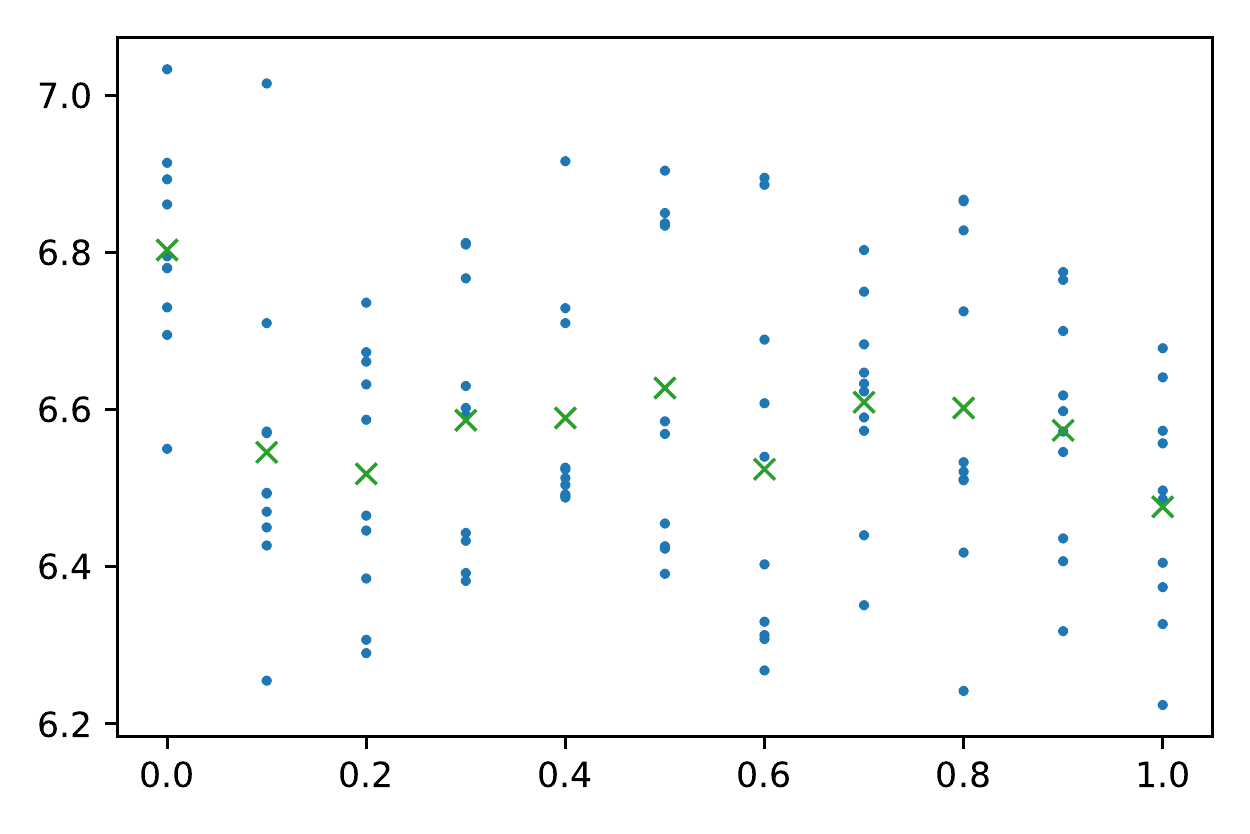}
\begin{picture}(0,0)
\put(-85,155){$\bar k$}
\put(105,25){$p$}
\end{picture}\vspace{-7mm}
         \caption{mean degree\vspace{3mm}}
         \label{fig:sqr_deg}
    \end{subfigure}
    \hfill
    \begin{subfigure}[b]{0.45\textwidth}
         \centering
          \includegraphics[width =\textwidth]{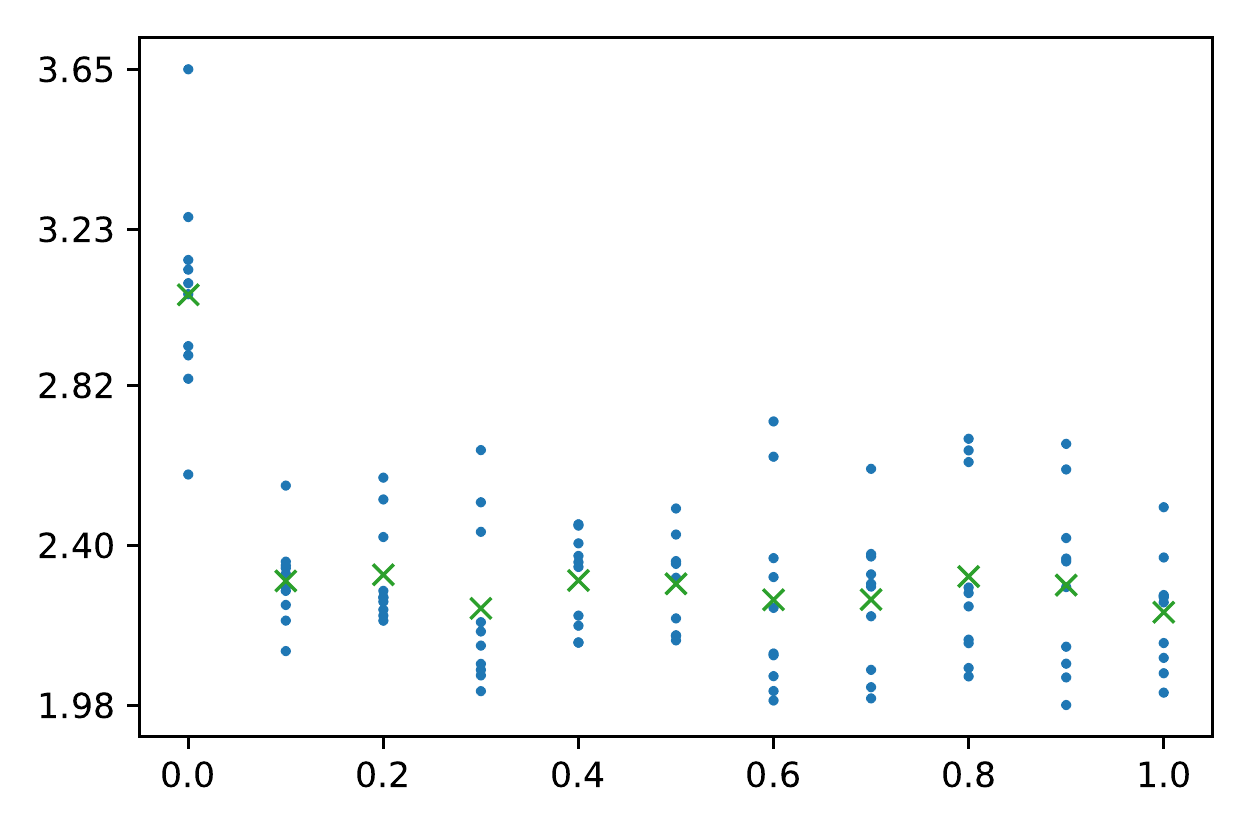}
\begin{picture}(0,0)
\put(-85,155){$\mbox{v}_k$}
\put(105,25){$p$}
\end{picture}\vspace{-7mm}
         \caption{degree variance\vspace{3mm}}
         \label{fig:sqr_var}
    \end{subfigure} 
    \\
    \begin{subfigure}[b]{0.45\textwidth}
         \centering
          \includegraphics[width =\textwidth]{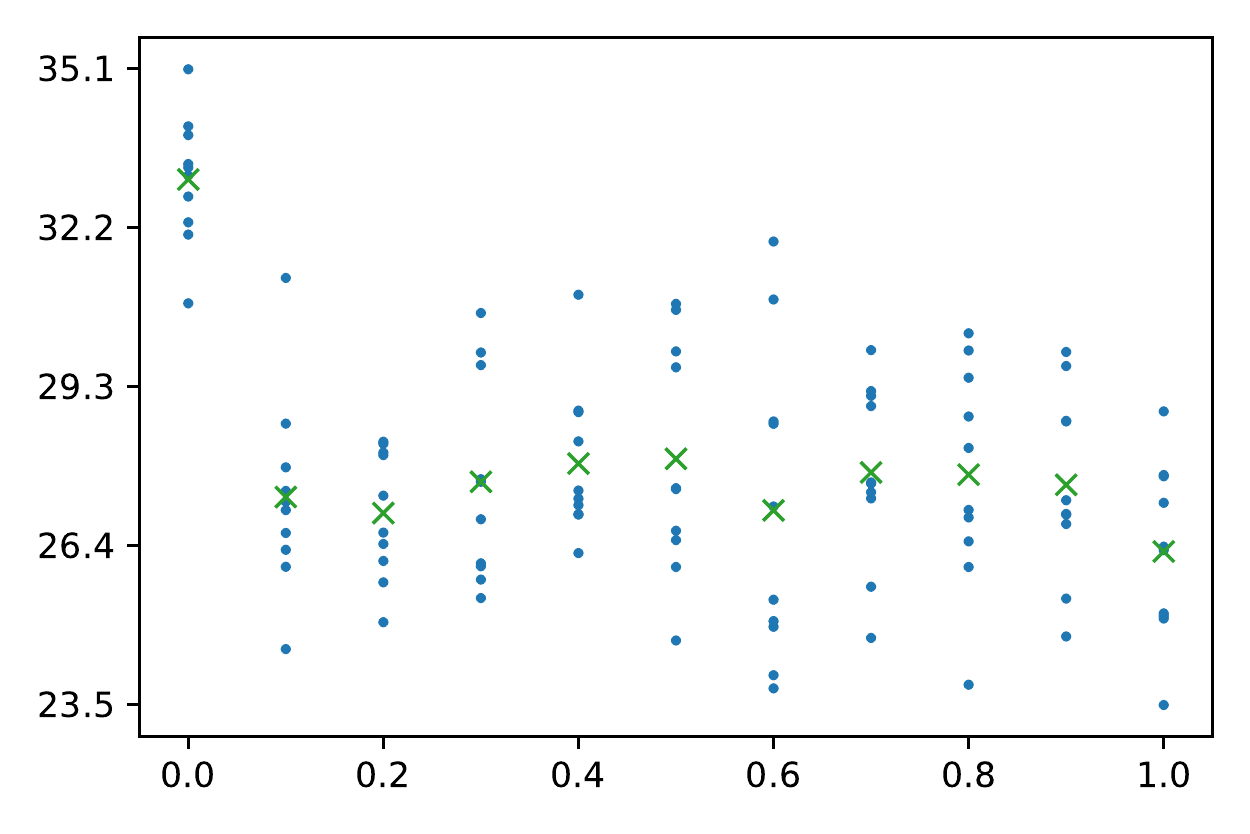}
\begin{picture}(0,0)
\put(-85,155){$\eta_\Box$}
\put(105,25){$p$}
\end{picture}\vspace{-7mm}
         \caption{percentage of squares\vspace{2mm}}
         \label{fig:sqr_sqr}
    \end{subfigure} 
    \hfill
    \begin{subfigure}[b]{0.45\textwidth}
         \centering
          \includegraphics[width =\textwidth]{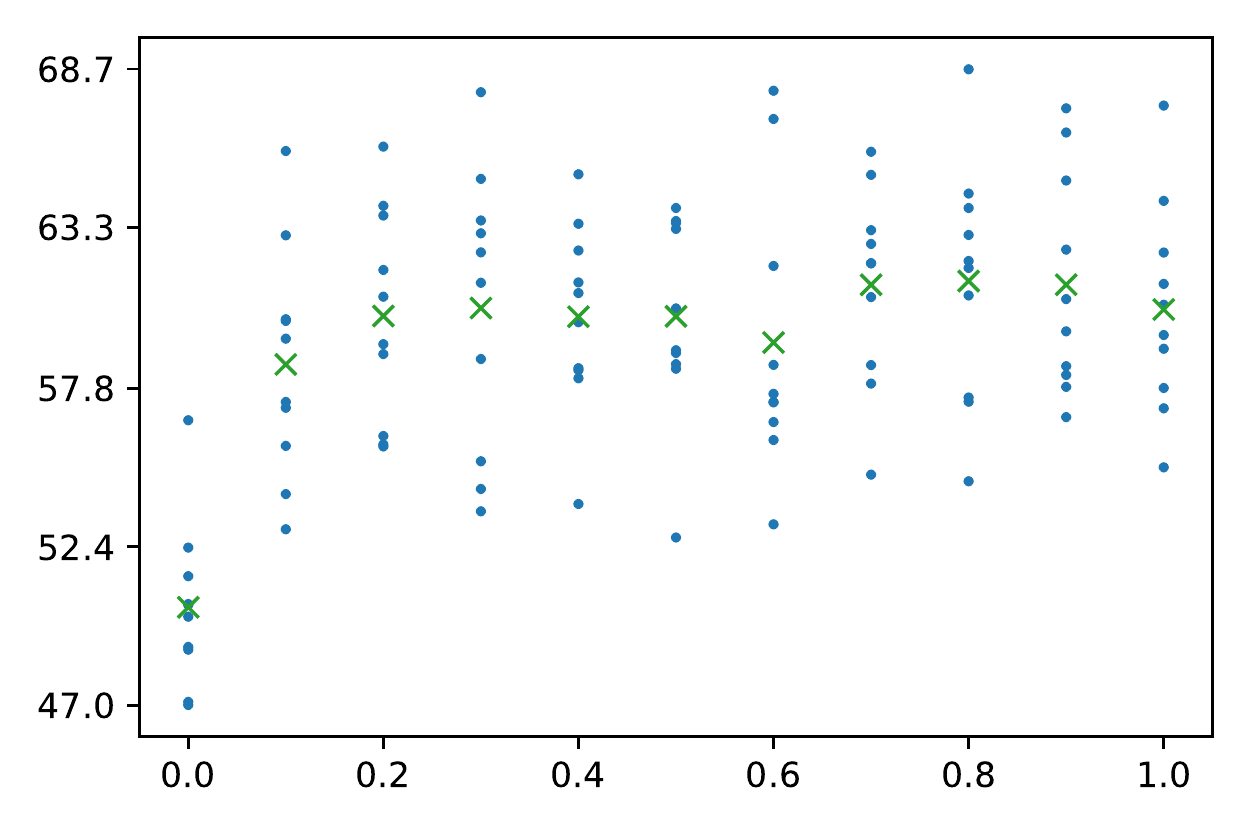}
\begin{picture}(0,0)
\put(-85,155){$\eta_{cube}$}
\put(105,25){$p$}
\end{picture}\vspace{-7mm}
         \caption{percentage of cubes\vspace{2mm}}
         \label{fig:sqr_cube}
    \end{subfigure}
    \caption{Outcomes of multiple simulations of the modified square model at $N=2000, \alpha=80, \beta=-20, \sigma=100$, with initial states in the form of Erd\H{o}s-R\'enyi graphs whose edge occupation probability $p$ is given on the horizontal axes. The green crosses indicate the mean of each vertical column.}
    \label{fig:sq_plots}
\end{figure}
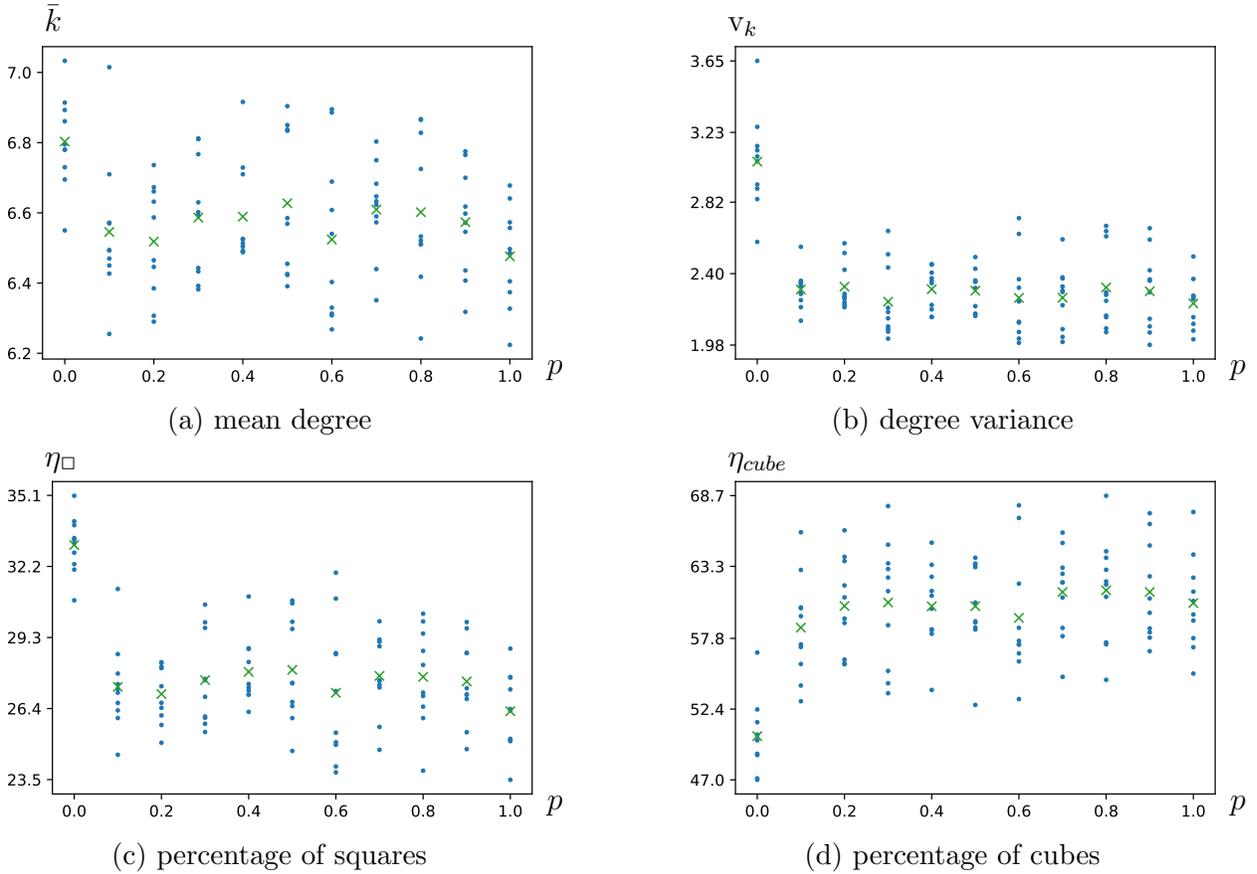

The phenomenology of the modified square model is similar to the previous section. If $\sigma$ is insufficiently large, one obtains graphs reminiscent of pure two-star model simulations, with very small numbers of geometric primitives. If $\sigma$ is large, one obtains graphs with extremely large numbers of squares and cubes, constituting a significant fraction of the maximal number possible at this degree. In this regime, there are jamming phenomena, similar to what has been described for the triangle model, in the sense that different Monte Carlo runs freeze at different numbers of geometric primitives, though for all runs the number of squares and cubes is very high, which represents a success of our model. 
We summarize the observational data of a particular set of runs in Fig.~\ref{fig:sq_plots}, which is analogous to Fig.~\ref{fig:tri_plots} for the triangle model. Some differences in fine details are that the degree variance is higher (given that the mean degree and percentage of squares are comparable to the mean degree and percentage of triangles in our simulations of the triangle model), while the cube production is extremely high.


\section{Preliminary analytic considerations}\label{sec: analytic}

While we have observed complex behaviors in our modified triangle and square models, such as emergence
of large counts of geometric primitives not explicitly coded into the graph Hamiltonian, as well as moderately jammed dynamics in the high $\sigma$ regime, we would like to emphasize that the graph Hamiltonians given by (\ref{2st}-\ref{Hamgen}) and either (\ref{Hsq}) or (\ref{Htr}) are rather simple in terms of their algebraic structure. In particular, if the modified square Hamiltonian (\ref{Hsq}), or the $\beta$-term of the two-star Hamiltonian (\ref{2st}), is expressed through the square of the adjacency matrix $q_{ij}$, they become a sum of terms each of which only depends on one single entry of $q$. The same holds for the $\al$-term of the two-start Hamiltonian as a function of $c_{ij}$. For the modified triangle Hamiltonian (\ref{Htr}), the same holds with respect to the matrix whose entries are $c_{ij}q_{ij}$.

This relative simplicity of the dependence on the adjacency matrix allows for a convenient rewriting of the partition functions of our models in terms of integrals, rather than discrete sums. A simpler version of the same procedure was applied in \cite{PN1} to the two-star model without any extra terms, resulting in a representation that elucidates the large $N$ behavior of the model. We shall give a demonstration below how to implement this sort of transformation for the modified square model. A similar (but more complicated) procedure must work for the modified triangle model. For the square model, in order to perform
the discrete summation over the components of $c$, one must first introduce the square of $c$ as an explicit summation variable. This is done as follows.

The partition function of the modified square model is 
\beq
Z(\al,\be,\s)=\sum_{\{c\}}\exp\Big[-H_{2*}(c)+\sigma\sum_{i<j} \de(q_{ij},2)\Big],
\label{Z}
\eeq
where $q$ is the square of the adjacency matrix given by (\ref{csq}), $\de$ is the Kronecker symbol (\ref{Kronecker}), and $H_{2*}$ is the two-star Hamiltonian (\ref{2st}). The sum $\sum_{\{c\}}$ is understood as a summation over the values 0 and 1 for each $c_{ij}$ with $i<j$. Whenever $c_{ii}$ occurs in any of our formulas, it is understood as 0, and $c_{ij}$ with $i>j$ is understood as $c_{ji}$. We shall be working with negative values of $\beta$ in $H_{2*}$, which guarantees convergence of various sums involved.
To introduce a summation over $q$ explicitly we insert in this formula the following representation of identity:
\beq
1=\sum_{\{q\}}\prod_{i<j} \de(q_{ij},\sum_k c_{ik}c_{kj})=\sum_{\{q\}}\int_0^{2\pi} \prod_{i<j}\frac{d\te_{ij}}{2\pi} \exp\Big[\iu\sum_{i<j} \te_{ij} (q_{ij}-\sum_k c_{ik}c_{kj})\Big],\label{1decomp}
\eeq
where $\sum_{\{q\}}$ is understood as a summation from 0 to $\infty$ over each $q_{ij}$ with $i<j$, and each integral over $\te_{ij}$ runs from 0 to $2\pi$.

It is convenient to rewrite the two-star Hamiltonian (\ref{2st}) as
\beqa
\dsty H_{2*}(c)&\dsty=-\al \sum_{ij} c_{ij} -\be \sum_{ikj} c_{ik}c_{kj}=-2\al \sum_{i<j} c_{ij} -\be \sum_{ik} c_{ik}^2-2\be\sum_{i<j}\sum_kc_{ik}c_{kj}\nonumber\\
&\dsty=-2(\al+\be)\sum_{i<j}c_{ij}-\be\sum_{i<j}\sum_kc_{ik}c_{kj}-\be\sum_{i<j}q_{ij}. \label{2stnew}
\eeqa
Once (\ref{1decomp}) and (\ref{2stnew}) have been substituted into (\ref{Z}), one gets
\beq
Z=\sum_{\{c,q\}}\int\limits_0^{2\pi} \prod_{i<j}\frac{d\te_{ij}}{2\pi} \exp\Big[2(\al+\be)\sum_{i<j}c_{ij}+\sum_{i<j} (\be-\iu\te_{ij})\sum_k c_{ik}c_{kj}+\sum_{i<j} \Big((\be+\iu\te_{ij}) q_{ij}+\s\de(q_{ij},2)\Big) \Big],
\eeq
The sum over $q$ can now be straightforwardly evaluated, independently for each $q_{ij}$, by applying
\beq
\sum_{q=0}^\infty e^{(\be+\iu\te)q+\sigma \de(q,2)}=\hspace{-5mm}\sum_{q=\{0,1,3,4,5,\ldots\}}\hspace{-5mm} e^{(\be+\iu\te)q}+e^{2(\be+\iu\te)+\s}=\frac{1}{1-e^{\be+\iu\te}}+e^{2(\be+\iu\te)}(e^\s-1)\equiv e^{{\cal F}(\te; \be,\s)},
\eeq
where the last equality serves as a definition of $\cal F$. One then gets
\beq
Z=\sum_{\{c\}}\int\limits_0^{2\pi} \prod_{i<j}\frac{d\te_{ij}}{2\pi} \exp\Big[2(\al+\be)\sum_{i<j}c_{ij}+\sum_{i<j}  {\cal F}(\te_{ij};\be,\s)\Big]\exp\Big[\sum_{i<j} (\be-\iu\te_{ij})\sum_k c_{ik}c_{kj}\Big].\label{Zqsum}
\eeq
The sum over $c_{ij}$ can be evaluated using a slightly more elaborate version of the Hubbard-Stratonovich trick applied in \cite{PN1} to the two-star model. Namely, for each given pair $(i,j)$, we write
\beqa
&\hspace{-2cm}\dsty\int \hspace{-1mm}\prod_{k=1}^N \frac{d\phi_{k}}{\sqrt{2\pi}} \exp\Big[(\be-\iu\te_{ij})\sum_k\Big(\frac12\phi_k\phi_k + \iu\phi_k (c_{ik}+c_{kj})\Big)\Big]=(\be-\iu\te_{ij})^{-N/2}e^{(\be-\iu\te_{ij})\sum_k (c_{ik}+c_{kj})^2/2}\hspace{-2cm}\nonumber\\
&\dsty=(\be-\iu\te_{ij})^{-N/2}\,e^{(\be-\iu\te_{ij})\sum_k (c_{ik}+c_{kj})/2}\,\exp\Big[(\be-\iu\te_{ij})\sum_k c_{ik}c_{kj}\Big].\label{HS}
\eeqa
One needs to introduce an independent vector $\phi_k$ of this sort for each pair $(i,j)$, giving a 3-index object $\phi_{ijk}$. Once this has been done, one can extract the last exponential of the last expression in (\ref{HS}) and substitute it into (\ref{Zqsum}). The argument of the exponential in the integrand is now a linear function of $c_{ij}$, and hence the sums over $c_{ij}$ are immediately evaluated, independently for different $(i,j)$, by applying the tautological formula
\beq
\sum_{c=0}^1 e^{ac}=1+e^a =\exp[\log(1+e^a)],
\eeq
as in \cite{PN1}.

As a result, one obtains a field-theoretic representation of $Z$ in terms of an integral over the phases $\te_{ij}$ and the real-valued fields $\phi_{ijk}$:
\beq
Z=\int\prod_{i<j}\left(\frac{d\te_{ij}}{2\pi}\prod_{k} \frac{d\phi_{ijk}}{\sqrt{2\pi}}\right) \exp\Big[\frac12\sum_{k;i<j}(\be-\iu\te_{ij})\phi_{ijk}\phi_{ijk}+ {\cal G}(\te_{ij},\phi_{ijk};\al,\be)+\sum_{i<j}  {\cal F}(\te_{ij};\be,\s)\Big],
\eeq
with
\beq
e^{\cal G}\equiv \prod_{i<j}(\be-\iu\te_{ij})^{N/2}\left[1+e^{2(\al+\be)-\sum_k \left\{\be-\iu(\te_{ik}+\te_{jk})/2+\iu(\be-\iu\te_{ik})\phi_{ikj}+\iu(\be-\iu\te_{jk})\phi_{jki}\right\}}\right],
\eeq
where it is understood that $\te_{ij}\equiv \te_{ji}$, $\phi_{ijk}\equiv\phi_{jik}$ and $\te_{ii}\equiv 0\equiv \phi_{iik}$.
 All the discrete summations are gone. This representation for $Z$ can be thought of as a reasonably conventional field theory on a complete graph, where $\te_{ij}$ are assigned to the graph edges, and $\phi_{ijk}$, to the 2-paths. The interactions between these fields are local in the sense that $\phi$ and $\te$ only interact with each other and among themselves if the geometric objects to which they are attached (edges and 2-paths) have common vertices. This form of locality leaves hope that some form of condensation will occur for $\phi$ and $\te$ at large $N$, which would open room for applications of $1/N$ analysis. While these topics would be interesting to explore, they go beyond our present treatment. We nonetheless feel that this field-theoretic representation for $Z$ is a welcome tool to keep in mind for the future.


\section{Comments on large-scale geometries}\label{sec: large_geo}

We have focused in our treatment on what can be called `small-scale geometry,' namely, on the emergence
of small geometric primitives out of an exponential random graph model that does not have the counts of these primitives explicitly appearing in its Hamiltonian. While we have not addressed them here explicitly,
in the back of our mind, we have kept the questions pertinent to `large-scale geometries,' such as those that
motivate the studies of \cite{Trugenberger,KT,KTB}, and more broadly, the subject of discrete quantum geometry/gravity \cite{QG,CDT,DGR}. Namely, one may ask whether the graphs one obtains from an exponential graph ensemble resemble, in an appropriate sense, discretized manifolds. We shall comment very briefly on the ingredients that should enter such studies.

In what sense a large graph may approximate a discretized manifold is in itself a rather subtle question.
It is not to be expected that a random graph, for example, one arising from an exponential ensemble, may
be regular enough on the scale of a single edge to match the rigid structure, say, of the edges of a simplicial complex. One expects that a sort of `smoothing procedure' must be applied first, akin to the block-spinning of lattice spin systems, so that an effective discrete geometry emerges at scales much bigger than a single edge but much smaller than the full graph. Defining this procedure must precede a thorough discussion of large-scale geometric properties of graphs.

At a more basic level, a few specific constructs exist in the random geometry literature that may be used to quantify the properties of graphs viewed as approximations to discretized manifolds. For example, one may consider the `volume profile' \cite{AJL,GS}, the number of vertices in a ball of a given size in terms of the graph distance viewed as a function of this distance. This quantity is expected to display various power laws depending on the scale (in relation to $N$), and is connected to the Hausdorff dimension \cite{durhuus}. One may also consider the spectral dimension, which arises from analysis of diffusion processes on graphs \cite{durhuus}. To efficiently employ these notions, however, one must have at one's disposal graphs much bigger than what straightforward Monte Carlo simulations allow (we have considered graphs with a few thousand vertices). Another important quantity that connects graphs and manifolds, frequently seen in the recent literature, is the Ollivier curvature \cite{ol1,ol2,ol3,ol4,ol5,ol6}. (The Ollivier curvature plays a central role in the random graph ensembles of \cite{Trugenberger,KT,KTB}, where it is used for the construction of the graph Hamiltonian, in analogy to how the Ricci curvature is used in the Einstein-Hilbert action. While we see the attraction of this approach, there does not have to be an {\it a priori} straightforward relation between the microscopic action defining a discrete theory and effective actions describing its continuum limits. For this reason, the main ingredient of \cite{Trugenberger,KT,KTB} that has entered our considerations is the hard-core condition that prevents graph collapse, rather than the graph Hamiltonians derived from the Ollivier curvature.)

\begin{figure}[t]
	\centering
    \begin{subfigure}[b]{0.45\textwidth}
	\includegraphics[width =\textwidth]{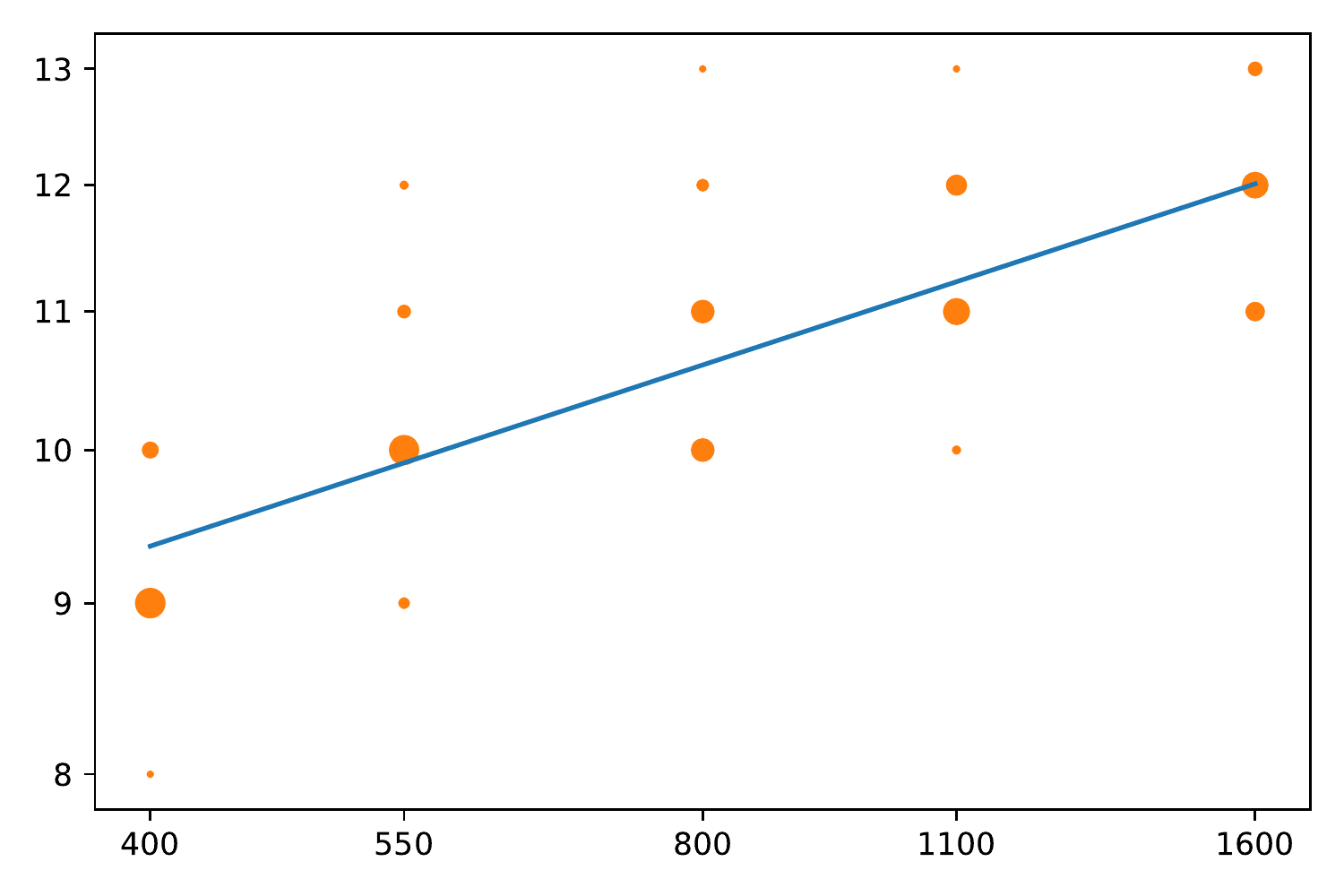}
\begin{picture}(0,0)
\put(20,135){$D$}
\put(192,33){$N$}
\end{picture}\vspace{-5mm}
    \end{subfigure}\hspace{10mm}
    \begin{subfigure}[b]{0.45\textwidth}
	\centering
	\includegraphics[width = \textwidth]{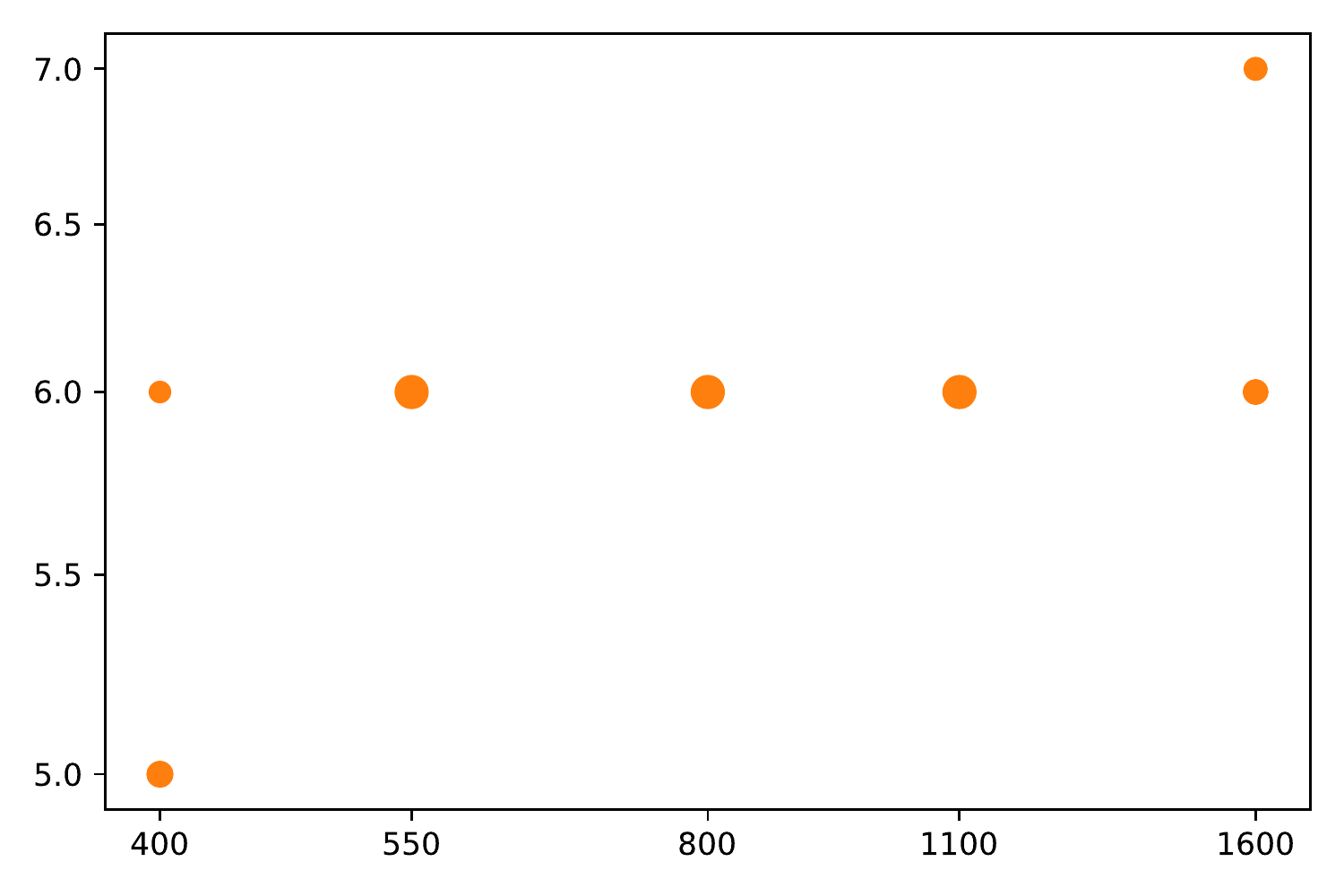}
\begin{picture}(0,0)
\put(-85,135){$D$}
\put(87,33){$N$}
\end{picture}\vspace{-5mm}
    \end{subfigure}
	\caption{Diameter values $D$ of the modified triangle model at $\alpha=177$, $\beta = -20$, $\sigma = 100$ (left) and the two-star model at $\alpha = 240$, $\beta = -20$ (right) plotted as a function of the number of vertices $N$ in a double-logarithmic presentation. At each graph size, 50 simulations have been performed. The area of the dots represents the number of hits this particular value of diameter has received in the course of the simulations. The blue line on the left plot is the best power-law fit that corresponds to the diameter varying as $N^{0.18}$.\vspace{-3mm}}
	\label{fig:diam_fit}
\end{figure}
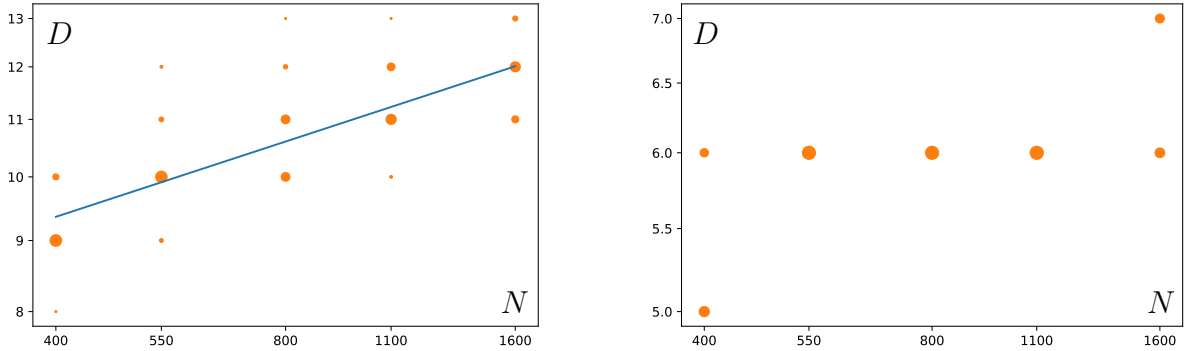

Perhaps the most basic quantifier of the geometric properties of graphs is the dependence of the graph diameter (the length of the shortest path between a given pair of vertices, maximized over the choice of this pair) on $N$, the number of vertices in the graph. This quantity would grow like a power of $N$ for regular lattices or regular simplicial complexes, but it is logarithmic in $N$ for sparse non-geometric random graphs, as the simplest Erd\H{o}s-R\'enyi model. (Some related questions have been considered in the random graph literature, such as the statistics of diameters of connected components \cite{diam1,diam2} in the Erd\H{o}s-R\'enyi model, and the shortest path statistics \cite{short1,short2,short3}.) It is unfortunately impossible to reliably distinguish a logarithm from power laws over the range of $N$ we can access, especially given the vertical spread of the data at any given $N$ (seen in Fig.~\ref{fig:diam_fit}) due to the jamming behavior combined with the slow variation of the diameter as a function of $N$. 
Simulating the model at considerably higher $N$ would require new ideas on numerical handling of exponential random graphs, and furthermore diameter evaluation becomes computationally costly for large graphs (it can be expressed through matrix multiplication and is therefore related to the numerical cost of multiplication of large matrices \cite{chung}; a particular approach is the Seidel algorithm \cite{seidel}). We can nonetheless assume that we are dealing with a power law and estimate the exponent by performing a linear regression fit (based on the entire cloud of observation points) on a double logarithmic scale. This fit is seen in Fig.~\ref{fig:diam_fit}, where we also show the much smaller diameter readings for the pure two-star model ($\s=0$) at mean degree 6. 
The specific power suggested by the fit is $N^{0.18 \pm 0.007}$.
Even apart from this estimate of the power-law exponent, the graphs our model produces have objectively very large diameters for an approximately regular graph of this size. Indeed, one can discuss for comparison diameters of random 6-regular graphs with 1600 vertices, where the analysis of \cite{diamreg} predicts a strict upper bound of at most 9, appreciably below the values we observe.
Note that, for random regular graphs, exceeding the bound of \cite{diamreg} by even one unit has a vanishing probability in the limit of a large number of vertices. Additionally, the end of \cite{diamreg} discusses plausible tighter bounds that give smaller values (not more than 7), consistent with what we actually observe for the two-star model (the right side of Fig.~\ref{fig:diam_fit}).


\section{Conclusions}\label{sec: conclusion}

We have designed two exponential random graph models obtained by modifying the conventional two-star random graph Hamiltonian \cite{PN1,AC} with either (\ref{Hsq}) or (\ref{Htr}). In an appropriate parameter regime where the new terms are given substantial weight, the models display pronounced geometric features, while avoiding the collapse or fracturing that plagues naive models with Hamiltonians based on subgraph counts, such as the Strauss model \cite{strauss,PN1,newexp}. We quantify the geometric features of our graphs by counting triangles, hexagons, squares and cubes. These motifs appear in very significant numbers, a sizable fraction of the maximal number possible in the relevant degree range. Such a behavior should be considered an emergent phenomenon, since the graph Hamiltonian explicitly depends only on properties of paths of length 2 and knows nothing about the bigger structures. By contrast, in the more familiar simpler models, as the Erd\H{o}s-R\'enyi or the two-star model, the probability of finding even one hexagon or cube in a large graph of a finite ($N$-independent) mean degree is simply zero. In the range of graph sizes we can access, the dependence of the graph diameter on the number of vertices is consistent with a power law, similarly to lattices and discretized spaces, but simulations of much larger graphs will be necessary to clearly distinguish this power law from a logarithmic behavior.

There is a number of interesting properties of our models that would deserve further exploration. Fig.~\ref{fig:tri_scan} suggest a phase transition separating geometric graphs that we have focused upon here, and non-geometric graphs reminiscent of the simple two-star model. It would be very interesting to elucidate the nature of this transition, and, more ambitiously, use it to construct a continuum limit of our models and connect them to the topics of random continuous geometry. All of these goals would require working with much bigger graphs, which is also the case for the topics of large-scale geometry outlined in the last section, and this remains a challenge. Similarly, it would be interesting to investigate further the analytic representation of our models in terms of real-valued fields on a complete graph, and connect them to topics of $1/N$ expansions. This would perhaps cast some further light on the nature of jamming visible on Figs.~\ref{fig:tri_plots} and \ref{fig:sq_plots}, which suggest a rich landscape of metastable configurations in our models.


\section*{Acknowledgments}

During the long course of formulation and development of this project, we have benefitted from conversations on related topics with Peter Grassberger, Christy Kelly, Dmitri Krioukov, Surachate Limkumnerd, Yong Lin, Yuki Sato and Bob Ziff. We especially thank Eytan Katzav, Luca Lionni and Carlo Trugenberger for discussions about geometry of random graphs and comments on the manuscript. Research of O.E. has been supported by the CUniverse research promotion initiative (CUAASC) at Chulalongkorn University. Research of T.C. has been supported by the Program Management Unit for Human Resources and Institutional Development, Research and Innovation (grant number B05F630108). Numerical simulations have been performed using the computational platforms of the Extreme Condition Physics Research Laboratory and Chula Intelligent and Complex Systems, Chulalongkorn University.
Graph visualizations are prepared using Gephi software \cite{gephi}. 

\section*{Notes added:}

\begin{enumerate}

\item Following the release of this article as a preprint, Alex Babeanu kindly brought to our attention an important earlier paper \cite{ChPl} that had previously escaped our attention. The mindset of this work is very similar to ours, and the authors construct a specific exponential random graph model adapted to producing manifold-like structures. The concrete Hamiltonian proposed there is, however, completely different from the ones used in this work, and has both advantages and disadvantages. In \cite{ChPl}, one first defines a `ring' of radius 1 around every vertex (the radius 1 neighborhood of this vertex with the original vertex itself excluded). Then the Hamiltonian is the difference of the diameter and radius of this ring (viewed as an isolated subgraph, not in terms of the original graph), summed over all choices of the center vertex. This definition of the Hamiltonian produces very high numbers of triangles, and graphs that are visibly manifold-like. It is, in this sense, superior to our considerations. This comes, however, with a significant price tag. First, while geometrically natural, the Hamiltonian of \cite{ChPl} does not have any simple expression through the adjacency matrix of the graph. Thus, it is perfectly suitable for numerical simulations, but holds relatively little promise for analytic work. Second, the Hamiltonian by design assigns probability zero whenever there is any vertex in the graph whose radius 1 ring (as defined above) is not connected. This very strong contraint is largely responsible for the manifold-like appearance of the random graphs of \cite{ChPl}, and avoiding such hard constraints has been one of the  key themes of our work. With all of this said, we view the model of \cite{ChPl} as an important contribution to the topic of emergence of geometry in random graphs, and it should certainly be kept in mind alongside with all other available models.

\item In a highly supportive and thought-provoking report, one of the anonymous referees has pointed out
the complementary perspective on our findings offered by the {\it graphon theory} \cite{graphon}. Studies of random graph models from the graphon perspective, in particular the edge-triangle Strauss model, can be found in \cite{Kr, NRS,RS,RKS}. These studies indicate systematic reasons
for the collapse behaviors in the edge-triangle model. Resolving these collapse behaviors and pushing the graphs closer to regular, manifold-like structures has been among the key motivations that have led us
to formulate the models studied in this article. While our focus has been on the sparse regime not directly related to the graphon theory, it is still instructive to review the nature of our graph Hamiltonians
as functions of the adjacency matrix (and furthermore, our models could equally well be studied in the dense regime where a finite fraction of all possible edges are filled). What has been pointed out by the referee is that, unlike the more conventional random graph models based on motif counts, our models do not have a natural graphon representation. This is due to the fact that our Hamiltonians do not vary smoothly with respect to the {\it cut metric} on the set of graphs, crucial for the graphon formulation. The models thereby avoid the general arguments for collapse in motif-based models developed in the graphon literature. This provides a deeper understanding for why the models developed here succeeded with the goals that motivated our considerations, while motif-based models do not, on account of their inherent collapse problems. 

\end{enumerate}


\end{document}